\documentclass[aps,prl,twocolumn,reprint,superscriptaddress]{revtex4}

\usepackage{hyperref}
\usepackage{graphicx}  
\usepackage{bm}        
\usepackage{amssymb}   
\usepackage{xspace}
\usepackage{verbatim}
\usepackage{color}
\usepackage{soul}
\usepackage{subfigure}
\usepackage[export]{adjustbox}
\usepackage{dcolumn}
\usepackage{amsthm,stmaryrd,mathtools}
\usepackage{txfonts}
\usepackage{braket}
\usepackage{amsmath}
\usepackage{xcolor}
\usepackage{array}
\usepackage{multirow}
\usepackage{tabularx}
\usepackage{booktabs}
\usepackage{lipsum}

\begin{document}

\title{Stark localization as a resource for weak-field sensing with super-Heisenberg precision}

\author{Xingjian He}%
\email{xjhe@std.uestc.edu.cn}
\affiliation{Institute of Fundamental and Frontier Sciences, University of Electronic Science and Technology of China, Chengdu 610051, China}

\author{Rozhin Yousefjani}%
\email{rozhinyousefjani@uestc.edu.cn}
\affiliation{Institute of Fundamental and Frontier Sciences, University of Electronic Science and Technology of China, Chengdu 610051, China}

\author{Abolfazl Bayat}%
\email{abolfazl.bayat@uestc.edu.cn}
\affiliation{Institute of Fundamental and Frontier Sciences, University of Electronic Science and Technology of China, Chengdu 610051, China}

\begin{abstract}
Gradient fields can effectively suppress particle tunneling in a lattice and localize the wave function at all energy scales, a phenomenon known as Stark localization. Here, we show that Stark systems can be used as a probe for the precise measurement of gradient fields, particularly in the weak-field regime where most sensors do not operate optimally. 
In the extended phase, Stark probes achieve super-Heisenberg precision, which is well beyond most of the known quantum sensing schemes. In the localized phase, the precision drops in a universal way showing fast convergence to the thermodynamic limit. For single-particle probes, we show that quantum-enhanced sensitivity, with super-Heisenberg precision, can be achieved through a simple position measurement for all the eigenstates across the entire spectrum. For such probes, we have identified several critical exponents of the Stark localization transition and established their relationship. Thermal fluctuations, whose universal behavior is identified, reduce the precision from super-Heisenberg to Heisenberg, still outperforming classical sensors. Multiparticle interacting probes also achieve super-Heisenberg scaling in their extended phase, which shows even further enhancement near the transition point. Quantum-enhanced sensitivity is still achievable even when state preparation time is included in resource analysis.  
\end{abstract}

\maketitle

\emph{Introduction.---} Cram\'{e}r-Rao inequality lies at the foundation of estimation theory~\cite{rao1992information,braunstein1994statistical,cramer1999mathematical}. It bounds the precision for estimating an unknown parameter $h$, quantified by
standard deviation $\delta h$, through  $\delta h \ge 1{/}\sqrt{\mathcal{M}\mathcal{F}}$, where $\mathcal{M}$ is the number of samples and $\mathcal{F}$ is Fisher information. Generally, Fisher information scales with the probe size $L$ as $\mathcal{F}\sim{L}^\beta$. While classical probes can at best achieve linear scaling, i.e., $\beta=1$ (standard limit), 
quantum features may enhance the precision to super-linear scaling with $\beta>1$~\cite{paris2009quantum,degen2017quantum,braun2018quantum}.
Originally, certain entangled states, known as GHZ states~\cite{greenberger1989going}, have been used for achieving $\beta=2$ (Heisenberg limit)~\cite{giovannetti2004quantum, leibfried2004toward,giovannetti2006quantum,banaszek2009quantum,giovannetti2011advances,frowis2011stable,demkowicz2012elusive,wang2018entanglement,kwon2019nonclassicality}. However, those schemes are prone to decoherence and are fundamentally limited to Heisenberg scaling (i.e., $\beta=2$). Alternatively, many-body probes can achieve quantum-enhanced sensitivity by harnessing a variety of quantum features. A class of such probes exploits various forms of criticality such as first-order~\cite{raghunandan2018high,heugel2019quantum,yang2019engineering}, second-order~\cite{zanardi2006ground,zanardi2007mixed,gu2008fidelity,zanardi2008quantum,invernizzi2008optimal,gu2010fidelity,gammelmark2011phase,skotiniotis2015quantum,rams2018limits,wei2019fidelity,chu2021dynamic,liu2021experimental,montenegro2021global,mirkhalaf2021criticality,di2021critical}, dissipative~\cite{baumann2010dicke,baden2014realization,klinder2015dynamical,rodriguez2017probing,fitzpatrick2017observation,fink2017observation,ilias2022criticality}, time crystals~\cite{montenegro2023quantum}, and topological~\cite{budich2020non,sarkar2022free,koch2022quantum,yu2022experimental} phase transitions. 
Other quantum many-body probes rely on quantum scars~\cite{dooley2021robust,desaules2021proposal,yoshinaga2022quantum,dooley2023entanglement} and Floquet driving~\cite{mishra2021driving,mishra2022integrable} as well as adaptive~\cite{wiseman1995adaptive,armen2002adaptive,fujiwara2006strong,higgins2007entanglement,berry2009perform,said2011nanoscale,okamoto2012experimental,bonato2016optimized,okamoto2017experimental,fernandez2017quantum}, continuous~\cite{gammelmark2014fisher,albarelli2017ultimate,rossi2020noisy,yang2022efficient,ilias2022criticality}, and sequential~\cite{burgarth2015quantum,montenegro2022sequential} measurements. 
There are two key open problems in many-body sensors. 
First, although the precision of these probes is not fundamentally bounded (i.e., no restriction on $\beta$), it is very hard to find quantum probes whose precision goes beyond Heisenberg sensitivity (i.e., $\beta>2$), with  few exceptions~\cite{boixo2007generalized,gu2008fidelity,roy2008exponentially,beau2017nonlinear,rams2018limits,wei2019fidelity}. Second, sensing weak fields with such probes is challenging as, for instance, critical points usually occur at finite field values, and the precision quickly drops away from that point.

The presence of a gradient field across a lattice makes the on-site energies off-resonant. Consequently, the tunneling rate is suppressed, and the wave function of the particles localizes in space. This is known as Stark localization~\cite{wannier1960wave} and has been exploited for inducing single-particle~\cite{fukuyama1973tightly,holthaus1995random,kolovsky2003bloch,kolovsky2008interplay,kolovsky2013wannier,van2019bloch} and many-body localization without disorder~\cite{van2019bloch,schulz2019stark,wu2019bath,bhakuni2020drive,bhakuni2020stability,yao2020many,chanda2020coexistence,taylor2020experimental,wang2021stark,zhang2021mobility,guo2021stark,yao2021many,doggen2022many,zisling2022transport,burin2022exact,bertoni2022local,lukin2022many,vernek2022robustness}, 
probing the geometry of nanostructures~\cite{pedersen2022stark}, protecting coherence~\cite{sarkar2022protecting}, investigating gauge theories~\cite{yang2020observation,halimeh2021gauge,zhou2022thermalization} and creating quantum scars~\cite{khemani2020localization,desaules2021proposal,kohlert2021experimental,scherg2021observing,doggen2021stark,su2022observation}. 
Stark localization has been experimentally observed in ion traps~\cite{morong2021observation}, optical lattices~\cite{preiss2015strongly,kohlert2021experimental}, and superconducting simulators~\cite{karamlou2022quantum}.
In the limit of large one-dimensional systems, in the absence of disorder and non-linearity, Stark localization takes place in infinitesimal fields for both single-particle~\cite{kolovsky2008interplay} and multiparticle interacting~\cite{doggen2021stark} cases.
One may wonder whether Stark systems can achieve quantum-enhanced precision sensing.
The key fact is that Stark localization transition takes place at the zero-field limit. Quantum-enhanced sensitivity in such a transition can be a breakthrough in ultra-precise sensing of weak fields, a domain in which most many-body sensors fail. 
In addition, since Stark localization happens across the whole spectrum it provides more thermal robustness than conventional criticality which is limited to the ground state.

In this letter, we show that Stark probes can achieve strong super-Heisenberg scaling in a region stretched all the way in the extended phase to  the transition point, for both single-particle and multiparticle interacting cases, allowing for ultraprecision weak-field sensing. 
In single-particle probes, one can achieve  $\beta=5.98$ for the ground state and $\beta=4.11$ for the mid-spectrum eigenstates, through a simple position measurement. 
Moreover, we determine several critical exponents and their universal relationship and show that thermal fluctuations reduce the precision from super-Heisenberg to Heisenberg scaling, still outperforming classical sensors. 
In addition, interacting many-body ground states achieve super-Heisenberg scaling, with $\beta=4.26$, which is close to the mid-spectrum single-particle case. This is because in the half-filling regime, the many-body ground state can be approximated by filling single-particle eigenstates up to midspectrum.

\emph{Ultimate precision limit.---} Let's consider a quantum probe whose density matrix $\rho(h)$ depends on an under scrutiny parameter $h$. 
To do so, one has to perform a measurement, described by a set of projective operators $\{\Pi_{i}\}$, on the probe. 
The result is described by a classical probability distribution in which each outcome appears with the probability  $p_i(h)=\mathrm{Tr}[\Pi_{i}\rho(h)]$. 
For this measurement setup, $\mathcal{F}$ in the Cram\'{e}r-Rao inequality is  called Classical Fisher Information (CFI), $\mathcal{F}_{C}(h)=\sum_{i}p_{i}(h)[\partial_{h}\ln p_{i}(h)]^2$~\cite{fisher1922mathematical}. 
Optimizing the CFI over all possible measurement setups leads to Quantum Fisher Information (QFI), namely $\mathcal{F}_Q(h)=\max_{\{\Pi_{i}\}}\mathcal{F}_C(h)$,  determining the ultimate precision limit achievable by a quantum probe~\cite{Meyer2021fisherinformationin}. A closed form for the QFI obtained through fidelity susceptibility $\chi(h)$~\cite{you2007fidelity} which for an incremental change in the parameter   
$\Delta h\rightarrow0$ can be approximated as $\chi(h){=}2[1-\tilde{F}(h)]/\Delta h^2$. Here, $\tilde{F}(h)=\mathrm{Tr}\sqrt{\rho(h)^{1/2}\rho(h+\Delta h)\rho(h)^{1/2}}$ is the fidelity between $\rho(h)$ and $\rho(h+\Delta h)$. 
The QFI then takes the form $\mathcal{F}_{Q}(h)=4\chi(h)$. 

\begin{figure}[t]
\includegraphics[width=\linewidth]{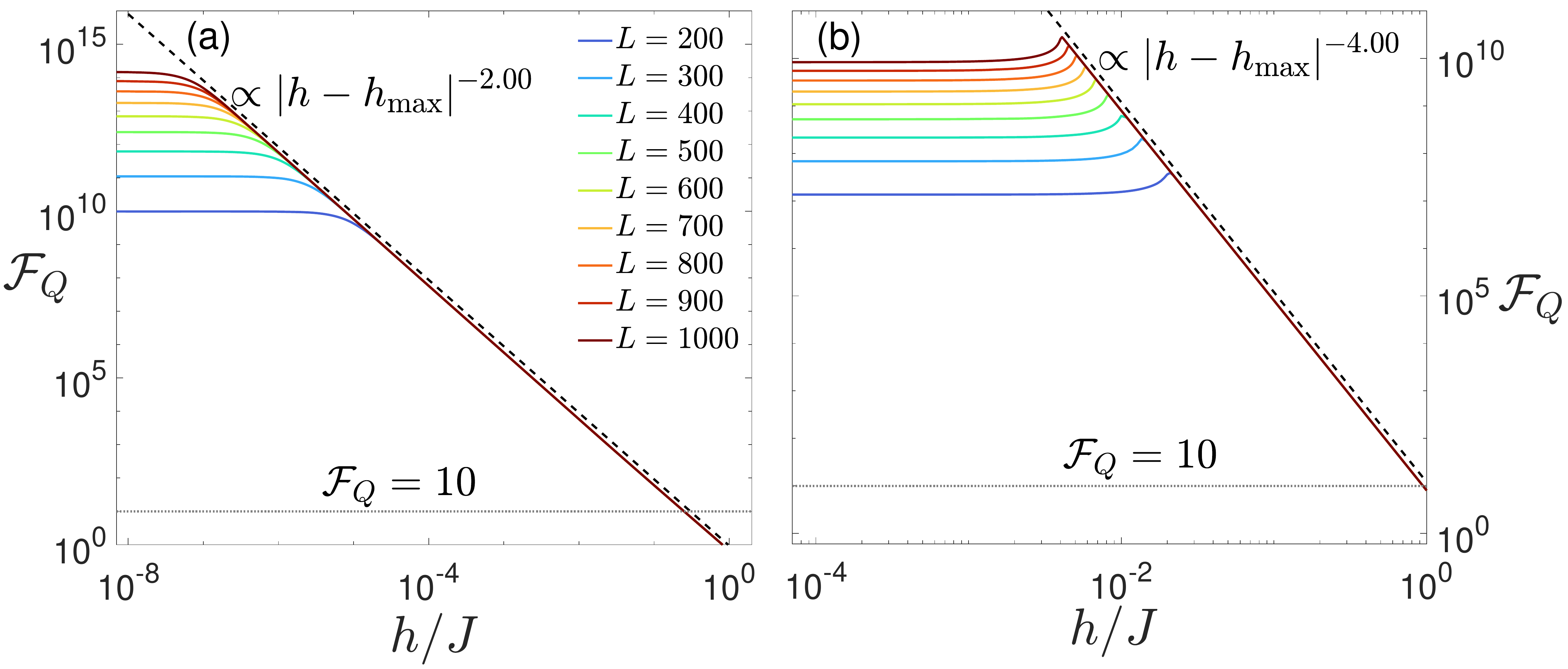} 
\includegraphics[width=\linewidth]{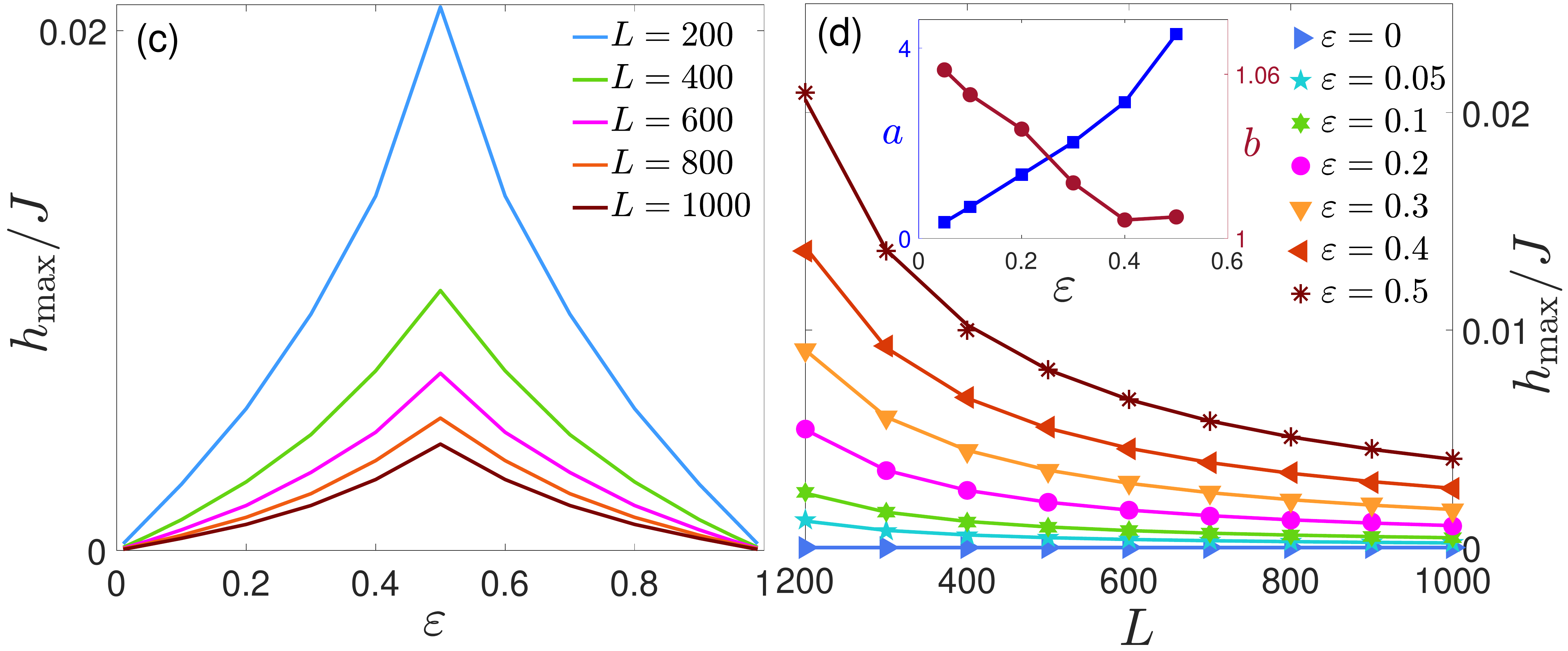}
\caption{The QFI versus $h/J$ when our probe with sizes $L$ is prepared in (a) the  ground state; and (b) the midspectrum eigenstate. The dashed lines in both panels are the best fit of $\mathcal{F}_Q$ in the localized phase, showing the behavior in the thermodynamic limit. (c) The transition point $h_{\max}/J$ versus energy $\varepsilon$, in various system sizes, indicating the emergence of mobility edges in finite systems. (d) The scaling of $h_{\max}/J$ with $L$ for various $\varepsilon$. Markers and solid lines represent numerical results and fitting function $h_{\max}=aL^{-b}$, respectively. Inset: $a$ and $b$ for different $\varepsilon$.
}\label{fig:Fig1}
\end{figure}

\emph{Single-particle probe.---}
We consider a one-dimensional probe with $L$ sites in which one particle can tunnel  to its neighbors with rate $J$. The probe is affected by a gradient field $h$ which we would like to estimate. 
The Hamiltonian is 
\begin{equation}\label{Eq:Stark_Hamiltonian}
H(h) = J\sum_{i=1}^{L-1} \left( \vert i\rangle\langle i+1\vert + \vert i+1\rangle\langle i\vert  \right) +  h \sum_{i=1}^{L}  i \vert i\rangle\langle i\vert.
\end{equation}
As $h/J$ increases, the system goes through a phase transition from an extended to a Stark localized phase~\cite{kolovsky2008interplay,van2019bloch,
schulz2019stark,chanda2020coexistence,yao2020many}. 
The transition is dramatic and affects the entire spectrum of the system which makes it very distinct from the conventional quantum phase transition as a ground state feature~\cite{zhang2021mobility}.

For our system in Eq.~(\ref{Eq:Stark_Hamiltonian}), in the thermodynamic limit (i.e., $L{\rightarrow}\infty$), the localization transition for all energy levels takes place at $h=h_c=0$~\cite{kolovsky2008interplay}. 
One can renormalize the energy of the system as $\varepsilon=(E-E_{\min})/(E_{\max}-E_{\min})$,
with $E_{\min}$ and $E_{\max}$ as extremal eigenenergies of the Hamiltonian $H(h)$ to fit the whole spectrum of the system within $0\le\varepsilon\le 1$. 
For any given $\varepsilon$, the QFI with respect to $h$ has been calculated for the closest eigenstate.
In Figs.~\ref{fig:Fig1}(a) and \ref{fig:Fig1}(b), we plot $\mathcal{F}_Q$ as a function of $h/J$ for various $L$, when our probe is in the ground state ($\varepsilon=0$) and a midspectrum eigenstate ($\varepsilon=0.5$), respectively.  
The QFI takes its maximum value at $h=h_{\max}$, which is expected to become the critical point, i.e., $h_{\max}=h_c$, in the thermodynamic limit. While for the ground state [see Fig.~\ref{fig:Fig1}(a)], the maximum takes place at vanishingly small fields, namely $h_{\max}\rightarrow0$. For the mid-spectrum [see Fig.~\ref{fig:Fig1}(b)], a clear peak for the QFI can be observed at non-zero $h_{\max}$.
Regardless of the energy levels, several features can be observed. First, by increasing $L$, the peak of the QFI, namely $\mathcal{F}_Q(h_{\max})$, dramatically enhances showing divergence in the thermodynamic limit. Second, the position of the peak gradually moves towards zero suggesting that in the thermodynamic limit one has $h_{\max}\rightarrow h_c=0$. Third, despite the decay of the QFI in the localized regime, its value remains high for a large interval of $h$, e.g., to have $\mathcal{F}_Q\ge 10$, the gradient field can be within the range $0\le h/J \le 0.25$ ($0\le h/J \le 1$) for the ground (midspectrum) state. Fourth, in the localized regime, after a certain threshold the QFI becomes size independent showing divergence to the thermodynamic limit, represented by dashed lines in Figs.~\ref{fig:Fig1}(a) and \ref{fig:Fig1}(b). 
Finite-size effects are evident in the initial plateaus of the QFI, representing the extended phase of the system. 
Interestingly, the dashed lines suggest an algebraic behavior of the QFI, namely $\mathcal{F}_{Q}(h)\propto|h-h_{\max}|^{-\alpha}$, in the localized phase which can be perfectly fitted  by  $\alpha=2.00$ for the ground state and $\alpha=4.00$ for the mid-spectrum eigenstates.
The fast convergence to the thermodynamic limit in the localized phase is discussed in more detail in the Supplementary Material.

To see how single-particle Stark localization transition depends on energy,  in Fig.~\ref{fig:Fig1}(c) we plot $h_{\max}$ as a function of $\varepsilon$ for various system sizes. The figure clearly indicates an energy-dependent transition which is known as mobility edge. 
In fact, midspectrum eigenstates are harder to localize than the eigenstates at the edges of the spectrum, a common feature that has also been observed in Anderson~\cite{girvin2019modern,anderson1958absence} and many-body~\cite{luitz2015many} localization. 
In the thermodynamic limit, mobility edge disappears, and the transition takes place at $h_{\max}=0$ across the whole spectrum.
To see how $h_{\max}$ decreases with an increasing system size at different energy scales, in Fig.~\ref{fig:Fig1}(d), we plot $h_{\max}$ versus $L$ for energies starting from the ground state ($\varepsilon=0$) to midspectrum ($\varepsilon=0.5$). 
Our numerical simulation (denoted by markers) is well described by the fitting function $h_{\max}(L)=aL^{-b}$ (solid lines). The exponents $a$ and $b$ are plotted as a function of $\varepsilon$ in the inset of Fig.~\ref{fig:Fig1}(d).
While the exponent $b$ shows a stable behavior around $b\simeq1$, the exponent $a$ changes dramatically from $a\simeq0$ (for the ground state) to $a\simeq4$ (for the midspectrum).

\begin{figure}[t]
\includegraphics[width=\linewidth]{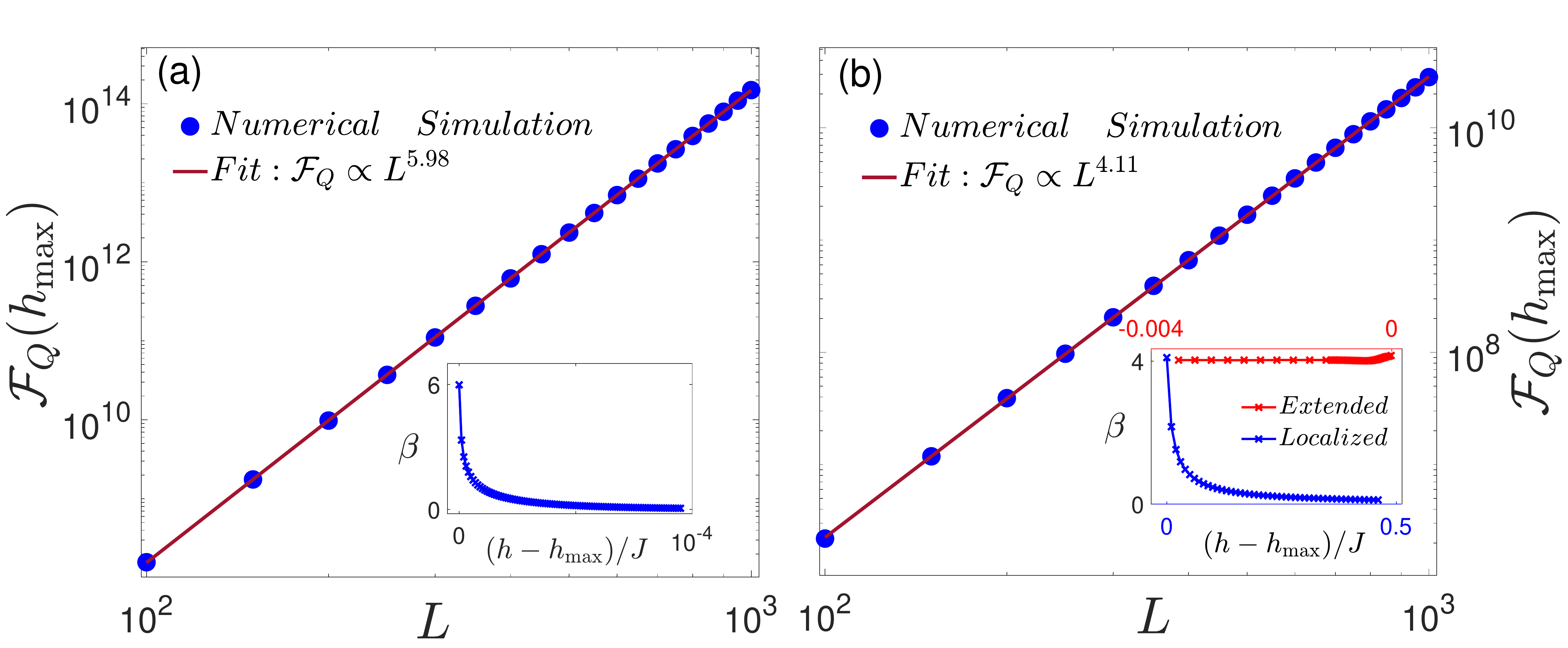}
\caption{The maximum of QFI (markers) as a function of $L$ for (a) the ground state and (b) the midspectrum eigenstate. 
The solid red lines are fitting of the form  $\mathcal{F}_{Q}(h_{\max})\propto L^{\beta}$ with $\beta=5.98$ for 
$\varepsilon=0$ and $\beta=4.11$ for $\varepsilon=0.5$. 
The inset of each panel represents the behavior of $\beta$  away from criticality. In the midspectrum (right panel), $\beta$ remains almost steady in the extended phase and drops in the localized phase.} \label{fig:Fig2}
\end{figure}

\emph{Super-Heisenberg sensitivity.---}
In the extended phase, the QFI heavily depends on size $L$.
To see how the QFI scales with the probe size, in Figs.~\ref{fig:Fig2}(a) and \ref{fig:Fig2}(b), we plot the QFI at the transition point, i.e., $\mathcal{F}_Q(h_{\max})$, as a function of $L$ for the ground ($\varepsilon=0$) and the midspectrum ($\varepsilon=0.5$) states, respectively. 
The QFI is shown by markers and red lines are fitting functions $\mathcal{F}_{Q}(h=h_{\max})\propto L^\beta$ with $\beta=5.98$ for the ground state and $\beta=4.11$ for the midspectrum states. 
This shows strong super-Heisenberg scaling in which $\beta$, by our knowledge, exceeds all other known many-body probes with local interaction and bounded spectrum.
We highlight these as our main results.
By  entering the localized regime  $\beta$  goes down and eventually vanishes. 
The smooth decay of $\beta$  is depicted in the insets of Figs.~\ref{fig:Fig2}(a) and ref{fig:Fig2}(b).

\begin{figure}[t]
\includegraphics[width=\linewidth]{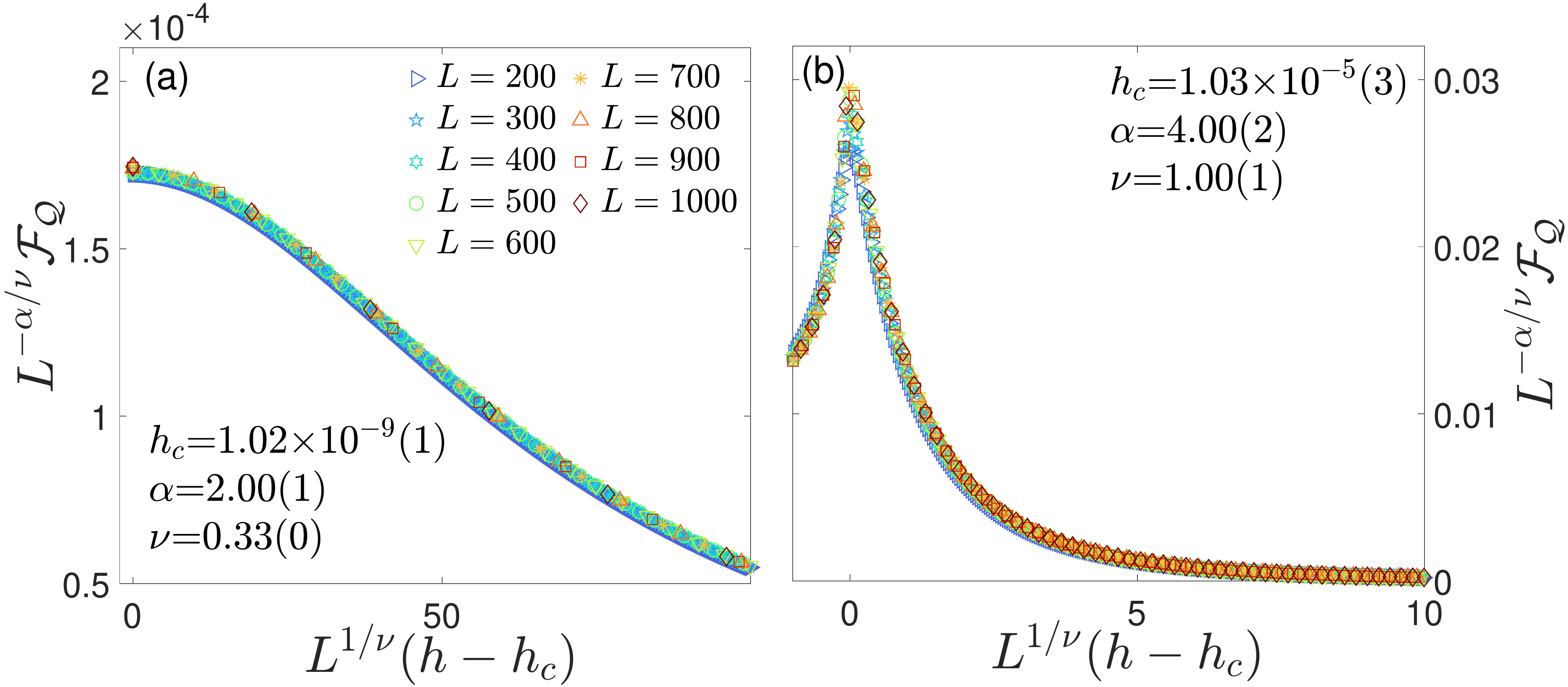} 
\caption{Finite-size scaling analysis using Eq.~(\ref{Eq.Finit-size-ansatz}) for (a) the ground state and (b) the midspectrum eigenstate. The optimal data collapse is obtained for the reported  $(h_c,\alpha,\nu)$. }\label{fig:Fig3}
\end{figure}

\emph{Finite-size scaling.---} 
On one side, Figs.~\ref{fig:Fig1}(a) and \ref{fig:Fig1}(b) suggest that in the thermodynamic limit (i.e., $L\rightarrow\infty$) QFI behaves as $\mathcal{F}_Q \propto |h-h_{\max}|^{-\alpha}$. On the other hand, the finite-size analysis at the transition point $h{=}h_{\max}$ [see Figs.~\ref{fig:Fig2}(a) and \ref{fig:Fig2}(b)] indicates $\mathcal{F}_{Q}(h=h_{\max})\propto L^\beta$. These two behaviors suggest 
\begin{equation}\label{Eq:Com-ansatz}
\mathcal{F}_{Q}(h) \propto \dfrac{1}{L^{-\beta} + A|h-h_{\max}|^{\alpha}},
\end{equation}
where $A$ is a constant. In the localized regime, where $|h-h_{\max}|^{\alpha} \gg L^{-\beta}$, the dependence on  $L$ becomes negligible. 
Thanks to the large value of $\beta$, the convergence to this limit is very rapid in our Stark probe; see the Supplementary Material for more details. 
The algebraic behavior of QFI hints that this transition might be of the second-order type.  
Any second-order phase transition is accompanied by a diverging length scale as $\xi\sim|h-h_c|^{-\nu}$, where the exponent $\nu$ controls the speed of divergence. 
In the case of localization transition, $\xi$ is indeed the localization length. 
To verify the nature of the transition and determine the critical exponents, we consider a conventional second-order finite-size scaling analysis. This implies that the QFI follows the ansatz
\begin{equation}\label{Eq.Finit-size-ansatz}
\mathcal{F}_{Q}(h)=L^{\alpha/\nu}g[L^{1/\nu}(h-h_c)],
\end{equation}
where, $g(\cdot)$ is an arbitrary function. To verify the above ansatz one can plot $L^{-\alpha{/}\nu}\mathcal{F}_Q(h)$, versus $L^{1{/}\nu}(h-h_c)$ for different $L$. 
By tuning the parameters $(h_c,\alpha,\nu)$ one tries to collapse all the curves of different system sizes on a single one.   
In Figs.~\ref{fig:Fig3}(a) and \ref{fig:Fig3}(b), we plot the best achievable data collapse, using the PYTHON package PYFSSA~\cite{andreas_sorge_2015_35293,melchert2009autoscale}, for $L=200$ to $L=1000$ for $\varepsilon=0$ and $\varepsilon=0.5$, respectively.
Our careful finite-size scaling analysis results in 
$(h_c,\alpha,\nu)=(1.02\times10^{-9}, 2.00, 0.33)$ and 
$(h_c,\alpha,\nu)=(1.03\times10^{-5}, 4.00, 1.00)$,
for the ground and the midspectrum states, respectively. The values obtained for $\alpha$ are fully consistent with the values extracted by independent data fitting in Figs.~\ref{fig:Fig1}(a) and \ref{fig:Fig1}(b).

The two Ans$\rm\ddot{a}$tze [Eq.~(\ref{Eq:Com-ansatz}) and Eq.~(\ref{Eq.Finit-size-ansatz})] describe the QFI and thus cannot be independent. Indeed, by factorizing $L^\beta$ from Eq.~(\ref{Eq:Com-ansatz}) one finds that these two will be the same if and only if 
\begin{equation}\label{Eq:exponents}
\beta=\dfrac{\alpha}{\nu}.
\end{equation}
This means that $\alpha$, $\beta$, and $\nu$ are not independent exponents, and two of them can describe the Stark transition.
In the Supplementary Material, by providing the values of $\alpha$, $\nu$, and $\beta$ we show the validity of Eq.~(\ref{Eq:exponents}) across the whole spectrum.

\emph{Optimal measurement.---}
Saturating the quantum Cram\'{e}r-Rao bound generally demands complex optimal measurements which may even depend on the unknown parameter $h$. 
Therefore, finding an experimentally feasible set of measurements with precision close to the Cram\'{e}r-Rao bound is highly desirable. 
Interestingly, in our probe, a simple position measurement described by local projective operators $\{\Pi_{i}{=}\vert i \rangle\langle i\vert\}_{i=1}^{L}$ saturates the QFI across the whole spectrum (see the Supplementary Material).

\begin{figure}[t]
\includegraphics[width=\linewidth]{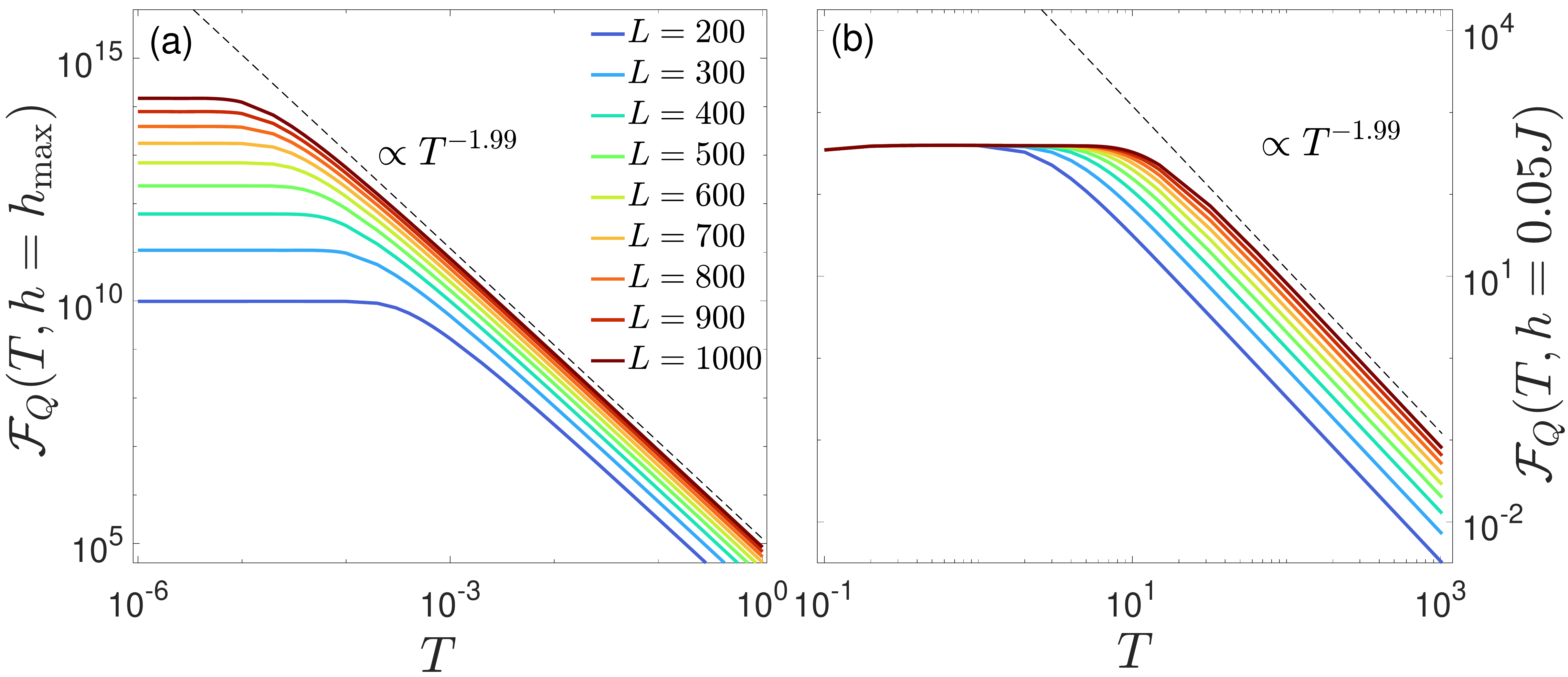}
\caption{The QFI of the thermal state as a function of temperature $T$ for various system sizes, when the probe is operating at (a) the transition point $h=h_{\max}=10^{-8}J$; and (b) the localized phase with $h=0.05J$.}
\label{fig:Fig4}
\end{figure}

\emph{Thermal probes.---} Apart from the ground state, accessing individual eigenstates is very difficult. In practice, our probe might be described by a thermal state $\rho(h,T){=}e^{-H/KT}/\mathrm{Tr}\left[ e^{-H/KT}\right]$, where $T$ is temperature, and $K$ is Boltzmann constant. 
Note that, the QFI is still computed with respect to $h$ and $T$ is only a parameter. 
We consider two different scenarios with $h$  in the extended phase, namely $h=10^{-8}J$, and in the localized phase, namely $h=0.05J$. 
The QFI, versus $h$, for both cases is plotted as a function of $T$ in Figs.~\ref{fig:Fig4}(a) and \ref{fig:Fig4}(b), respectively. The QFI starts with a plateau, whose width is $KT\simeq \Delta E$, where $\Delta E=E_2-E_1$ is the energy gap. In this regime, the thermal state is  described by the ground state. That is why in the extended phase [Fig.~\ref{fig:Fig4}(a)], the plateaus are separated for each $L$ while in the localized phase [Fig.~\ref{fig:Fig4}(b)], they collapse. Interesting behavior can be observed for $KT>\Delta E$, where the QFI decays as $T$ increases. In this regime, as shown in Fig.~\ref{fig:Fig4}, for both localized and extended probes, the QFI decays as $\mathcal{F}_Q\propto c(L) T^{-\mu}$, where $c(L)\propto L^\gamma$, with $\gamma=2.0$ (see the Supplementary Material) and $\mu=1.99$ is a universal exponent. Therefore, in the limit of $KT>\Delta E$ the QFI is universally described as
\begin{equation}
\mathcal{F}_Q(h)\sim f(h) T^{-\mu} L^{\gamma},
\end{equation}
which shows Heisenberg scaling with respect to system size, outperforming classical sensors. 

\begin{figure}[t!]
\includegraphics[width=\linewidth]{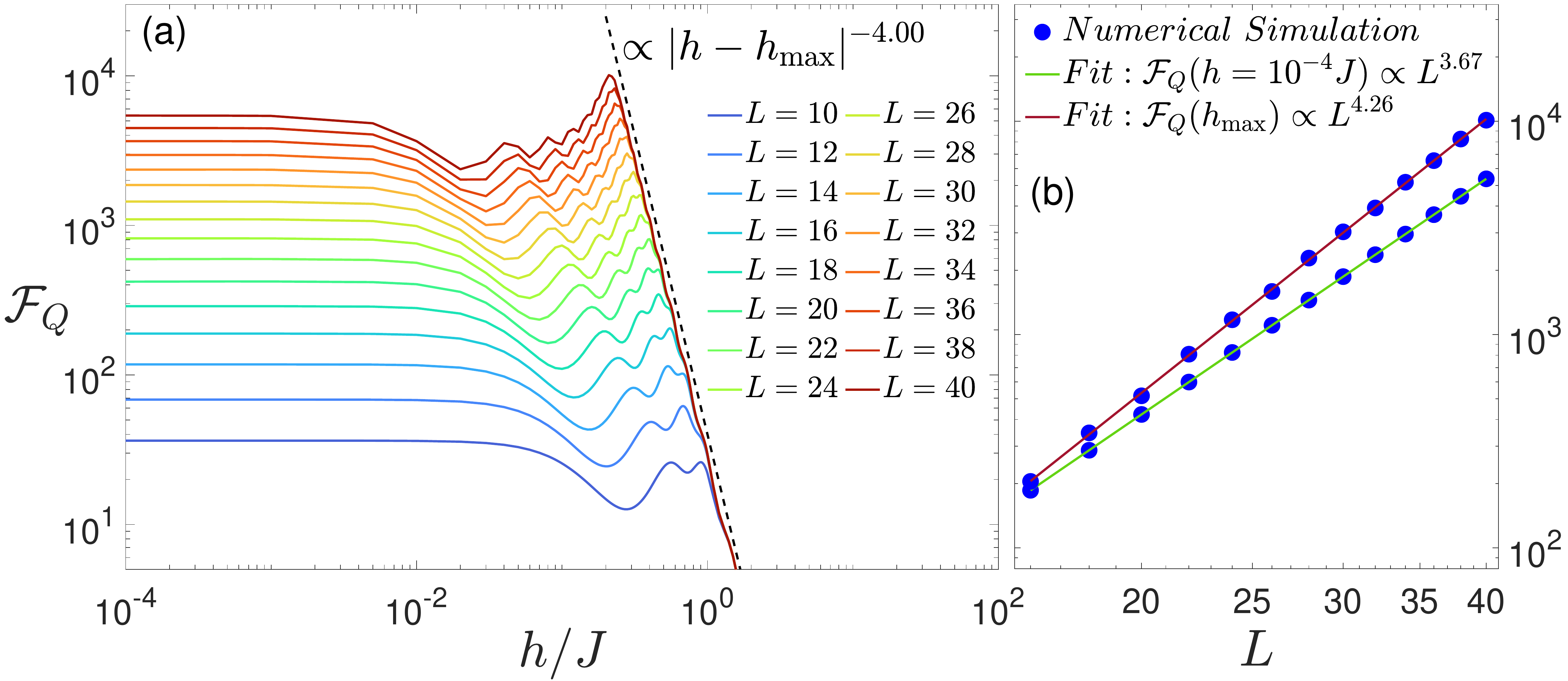}
\caption{(a) The QFI of a many-body interacting system with $N=L{/}2$ particles versus $h/J$ when the probe of size $L$ is in the ground state. 
The dashed line is the best fitting function of $\vert h-h_{\max}\vert$ in the localized phase illustrating the system's behavior in the thermodynamic limit.
(b) The  QFI at the extended and transition point, namely $h=h_{\max}$, as a function of $L$. }\label{fig:Fig5}
\end{figure}

\emph{Many-body interacting probes.---} In this section, we show that our probe can also operate in the case of multiparticle interacting systems. We consider a one-dimensional probe of size $L$ with $N{=}L{/}2$ particles interacting via Hamiltonian 
\begin{equation}
H(h) = J\sum_{i=1}^{L-1}(\sigma_{i}^{x}\sigma_{i+1}^{x}+\sigma_{i}^{y}\sigma_{i+1}^{y}+\sigma_{i}^{z}\sigma_{i+1}^{z}) + h\sum_{i=1}^{L} i \sigma_{i}^{z},
\end{equation}
where $\sigma_{i}^{x,y,z}$ are the Pauli operators and $h$, again, is the strength of the gradient field.  
To benefit from Matrix Product State (MPS) analysis for capturing large system sizes, we only focus on the ground state. 
In Fig.~\ref{fig:Fig5}(a) we plot $\mathcal{F}_{Q}$ as a function of $h/J$, in a half-filling regime, for various $L$. We use 
exact diagonalization for systems up to $L=20$ and MPS (using TeNPy Library~\cite{TENPY}), with bond dimension $\chi=1000$,  for larger $L$'s.
Surprisingly, the calculated QFI in the interacting many-body system behaves qualitatively similar to the single-particle case in the midspectrum [see Fig.~\ref{fig:Fig1}(b)].
While it shows a steady behavior in the extended phase, it fluctuates and eventually peaks at a non-zero $h_{\max}$.
Clearly, $\mathcal{F}_{Q}(h_{\max})$ enhances by increasing $L$, signaling the divergence in the thermodynamic limit. 
In the localized phase, the QFI decays algebraically as $\mathcal{F}_{Q}\propto\vert h-h_{\max}\vert^{-\alpha}$ with $\alpha=4.00$.
Implementing careful finite-size scaling analysis results in $(h_c,\alpha,\nu)=(1.16\times10^{-5},4.03,1.02)$, showing an agreement with the single-particle probes prepared in the midspectrum [see Fig.~\ref{fig:Fig1}(b)].

To see the scaling of QFI, in Fig.~\ref{fig:Fig5}(b) we plot $\mathcal{F}_{Q}(h_{\max})$ as a function of $L$ which shows algebraic scaling as $\mathcal{F}_{Q}(h_{\max})\propto L^{\beta}$, with $\beta=3.67$ in the extended phase which is further enhanced at the transition point to  $\beta=4.26$.  
The exponent, in particular near the transition point, is also close to the one from the midspectrum single-particle probes. 
This interesting observation can be described in a hand-waving way. The ground state of a half-filling system, at least in the \emph{noninteracting} limit, can be constructed by filling single-particle eigenstates up to mid-spectrum. 
Although the presence of interaction makes this picture less precise, it shows that the overall sensing power of the probe is dominantly defined by midspectrum single-particle eigenstates. The clear distinction between the curves in Fig.~\ref{fig:Fig1}(b) (i.e., midspectrum of single-particle probe) and Fig.~\ref{fig:Fig5}(a) (i.e., ground state of many-body probe) is also indicated by a small deviation between their exponent $\beta$, which can be assigned to the effect of interaction.    
Note that the gradual movement of $h_{\max}$ towards zero suggests that, similar to the single-particle case, the Stark many-body localization transition also takes place at infinitesimal values of the field. This is consistent with the results of Ref.~\cite{doggen2021stark}, found for Stark localization transition in the absence of disorder and nonlinearity, and shows that many-body Stark probes also serve for weak-field sensing with super-Heisenberg precision.

\emph{Resource analysis.---} If probe size $L$ is the only resource that we care about, then the QFI is the best figure of merit for resource efficiency. However, eigenstate preparation (e.g., ground state) is time-consuming which might be considered as another resource. For instance, using adiabatic evolution for ground state initialization near criticality requires a size-dependent time which scales as $t\sim L^z$. The critical exponent $z$ quantifies the energy gap closing at the criticality $\Delta E\sim L^{-z}$~\cite{rams2018limits}. To include time as a resource, one may consider normalized QFI  $\mathcal{F}_Q/t$ as the figure of merit~\cite{rams2018limits,chu2021dynamic,montenegro2022sequential} which turns out to be  $\mathcal{F}_Q/t{\sim} L^{\beta-z}$. As discussed in the Supplementary Material, we have estimated $z\simeq2$, resulting in $\mathcal{F}_Q/t\sim L^4$, for the single-particle and $z\simeq0.81$, resulting in $\mathcal{F}_Q/t\sim L^{3.45}$, for the many-body probes which both still show strong quantum-enhanced precision.

\emph{Conclusion.---}  
We have shown that Stark probes are extremely precise in measuring weak gradient fields, achieving strong super-Heisenberg precision over a region which stretches all over the extended phase  to the transition point. Quantum-enhanced sensitivity is still achievable even when preparation time is included in the resource analysis. We have considered both single-particle and many-body interacting systems. In the case of single particles, we have shown that super-Heisenberg precision can be achieved through a simple position measurement across the whole spectrum. In this regime, we have determined the critical exponents and their relationship. We have also identified the universal behavior of the probe at thermal equilibrium, which shows that as temperature increases the scaling of precision reduces from super-Heisenberg to Heisenberg. For interacting many-body ground state, while a strong super-Heisenberg precision can still be achieved, the scaling exponents become close to the midspectrum single-particle case. This can be understood as the many-body ground state can be approximately constructed by filling single-particle eigenstates until midspectrum.

\emph{Acknowledgment.--} A.B. acknowledges support from the National Key R\&D Program of China (Grant No. 2018YFA0306703), the National Science Foundation of China (Grants No. 12050410253, No. 92065115, and No. 12274059), and the Ministry of Science and Technology of China (Grant No. QNJ2021167001L). R.Y. thanks the National Science Foundation of China for the International Young Scientists Fund (Grant No. 12250410242).


\begin{thebibliography}{116}
\expandafter\ifx\csname natexlab\endcsname\relax\def\natexlab#1{#1}\fi
\expandafter\ifx\csname bibnamefont\endcsname\relax
  \def\bibnamefont#1{#1}\fi
\expandafter\ifx\csname bibfnamefont\endcsname\relax
  \def\bibfnamefont#1{#1}\fi
\expandafter\ifx\csname citenamefont\endcsname\relax
  \def\citenamefont#1{#1}\fi
\expandafter\ifx\csname url\endcsname\relax
  \def\url#1{\texttt{#1}}\fi
\expandafter\ifx\csname urlprefix\endcsname\relax\def\urlprefix{URL }\fi
\providecommand{\bibinfo}[2]{#2}
\providecommand{\eprint}[2][]{\url{#2}}

\bibitem[{\citenamefont{Rao}(1992)}]{rao1992information}
\bibinfo{author}{\bibfnamefont{C.~R.} \bibnamefont{Rao}}, in
  \emph{\bibinfo{booktitle}{Breakthroughs in statistics}}
  (\bibinfo{publisher}{Springer, New York}, \bibinfo{year}{1992}), pp.
  \bibinfo{pages}{235--247}.

\bibitem[{\citenamefont{Braunstein and
  Caves}(1994)}]{braunstein1994statistical}
\bibinfo{author}{\bibfnamefont{S.~L.} \bibnamefont{Braunstein}}
  \bibnamefont{and} \bibinfo{author}{\bibfnamefont{C.~M.} \bibnamefont{Caves}},
  \bibinfo{journal}{Phys. Rev. Lett.} \textbf{\bibinfo{volume}{72}},
  \bibinfo{pages}{3439} (\bibinfo{year}{1994}).

\bibitem[{\citenamefont{Cram{\'e}r}(1999)}]{cramer1999mathematical}
\bibinfo{author}{\bibfnamefont{H.}~\bibnamefont{Cram{\'e}r}},
  \emph{\bibinfo{title}{Mathematical methods of statistics}},
   (\bibinfo{publisher}{Princeton University Press, Princeton, NJ},
  \bibinfo{year}{1999}), Vol.~\bibinfo{volume}{43}.

\bibitem[{\citenamefont{Paris}(2009)}]{paris2009quantum}
\bibinfo{author}{\bibfnamefont{M.~G.} \bibnamefont{Paris}},
  \bibinfo{journal}{Int. J. Quantum Inf.} \textbf{\bibinfo{volume}{07}},
  \bibinfo{pages}{125} (\bibinfo{year}{2009}).

\bibitem[{\citenamefont{Degen et~al.}(2017)\citenamefont{Degen, Reinhard, and
  Cappellaro}}]{degen2017quantum}
\bibinfo{author}{\bibfnamefont{C.~L.} \bibnamefont{Degen}},
  \bibinfo{author}{\bibfnamefont{F.}~\bibnamefont{Reinhard}}, \bibnamefont{and}
  \bibinfo{author}{\bibfnamefont{P.}~\bibnamefont{Cappellaro}},
  \bibinfo{journal}{Rev. Mod. Phys.} \textbf{\bibinfo{volume}{89}},
  \bibinfo{pages}{035002} (\bibinfo{year}{2017}).

\bibitem[{\citenamefont{Braun et~al.}(2018)\citenamefont{Braun, Adesso,
  Benatti, Floreanini, Marzolino, Mitchell, and Pirandola}}]{braun2018quantum}
\bibinfo{author}{\bibfnamefont{D.}~\bibnamefont{Braun}},
  \bibinfo{author}{\bibfnamefont{G.}~\bibnamefont{Adesso}},
  \bibinfo{author}{\bibfnamefont{F.}~\bibnamefont{Benatti}},
  \bibinfo{author}{\bibfnamefont{R.}~\bibnamefont{Floreanini}},
  \bibinfo{author}{\bibfnamefont{U.}~\bibnamefont{Marzolino}},
  \bibinfo{author}{\bibfnamefont{M.~W.} \bibnamefont{Mitchell}},
  \bibnamefont{and}
  \bibinfo{author}{\bibfnamefont{S.}~\bibnamefont{Pirandola}},
  \bibinfo{journal}{Rev. Mod. Phys.} \textbf{\bibinfo{volume}{90}},
  \bibinfo{pages}{035006} (\bibinfo{year}{2018}).

\bibitem[{\citenamefont{Greenberger et~al.}(1989)\citenamefont{Greenberger,
  Horne, and Zeilinger}}]{greenberger1989going}
\bibinfo{author}{\bibfnamefont{D.~M.} \bibnamefont{Greenberger}},
  \bibinfo{author}{\bibfnamefont{M.~A.} \bibnamefont{Horne}}, \bibnamefont{and}
  \bibinfo{author}{\bibfnamefont{A.}~\bibnamefont{Zeilinger}}, in
  \emph{\bibinfo{booktitle}{Bell’s Theorem, Quantum Theory and Conceptions of
  the Universe}} (\bibinfo{publisher}{Springer, New York}, \bibinfo{year}{1989}), pp.
  \bibinfo{pages}{69--72}.

\bibitem[{\citenamefont{Giovannetti et~al.}(2004)\citenamefont{Giovannetti,
  Lloyd, and Maccone}}]{giovannetti2004quantum}
\bibinfo{author}{\bibfnamefont{V.}~\bibnamefont{Giovannetti}},
  \bibinfo{author}{\bibfnamefont{S.}~\bibnamefont{Lloyd}}, \bibnamefont{and}
  \bibinfo{author}{\bibfnamefont{L.}~\bibnamefont{Maccone}},
  \bibinfo{journal}{Science} \textbf{\bibinfo{volume}{306}},
  \bibinfo{pages}{1330} (\bibinfo{year}{2004}).

\bibitem[{\citenamefont{Leibfried et~al.}(2004)\citenamefont{Leibfried,
  Barrett, Schaetz, Britton, Chiaverini, Itano, Jost, Langer, and
  Wineland}}]{leibfried2004toward}
\bibinfo{author}{\bibfnamefont{D.}~\bibnamefont{Leibfried}},
  \bibinfo{author}{\bibfnamefont{M.~D.} \bibnamefont{Barrett}},
  \bibinfo{author}{\bibfnamefont{T.}~\bibnamefont{Schaetz}},
  \bibinfo{author}{\bibfnamefont{J.}~\bibnamefont{Britton}},
  \bibinfo{author}{\bibfnamefont{J.}~\bibnamefont{Chiaverini}},
  \bibinfo{author}{\bibfnamefont{W.~M.} \bibnamefont{Itano}},
  \bibinfo{author}{\bibfnamefont{J.~D.} \bibnamefont{Jost}},
  \bibinfo{author}{\bibfnamefont{C.}~\bibnamefont{Langer}}, \bibnamefont{and}
  \bibinfo{author}{\bibfnamefont{D.~J.} \bibnamefont{Wineland}},
  \bibinfo{journal}{Science} \textbf{\bibinfo{volume}{304}},
  \bibinfo{pages}{1476} (\bibinfo{year}{2004}).

\bibitem[{\citenamefont{Giovannetti et~al.}(2006)\citenamefont{Giovannetti,
  Lloyd, and Maccone}}]{giovannetti2006quantum}
\bibinfo{author}{\bibfnamefont{V.}~\bibnamefont{Giovannetti}},
  \bibinfo{author}{\bibfnamefont{S.}~\bibnamefont{Lloyd}}, \bibnamefont{and}
  \bibinfo{author}{\bibfnamefont{L.}~\bibnamefont{Maccone}},
  \bibinfo{journal}{Phys. Rev. Lett.} \textbf{\bibinfo{volume}{96}},
  \bibinfo{pages}{010401} (\bibinfo{year}{2006}).

\bibitem[{\citenamefont{Banaszek et~al.}(2009)\citenamefont{Banaszek,
  Demkowicz-Dobrza{\'n}ski, and Walmsley}}]{banaszek2009quantum}
\bibinfo{author}{\bibfnamefont{K.}~\bibnamefont{Banaszek}},
  \bibinfo{author}{\bibfnamefont{R.}~\bibnamefont{Demkowicz-Dobrza{\'n}ski}},
  \bibnamefont{and} \bibinfo{author}{\bibfnamefont{I.~A.}
  \bibnamefont{Walmsley}}, \bibinfo{journal}{Nat. Photonics}
  \textbf{\bibinfo{volume}{3}}, \bibinfo{pages}{673} (\bibinfo{year}{2009}).

\bibitem[{\citenamefont{Giovannetti et~al.}(2011)\citenamefont{Giovannetti,
  Lloyd, and Maccone}}]{giovannetti2011advances}
\bibinfo{author}{\bibfnamefont{V.}~\bibnamefont{Giovannetti}},
  \bibinfo{author}{\bibfnamefont{S.}~\bibnamefont{Lloyd}}, \bibnamefont{and}
  \bibinfo{author}{\bibfnamefont{L.}~\bibnamefont{Maccone}},
  \bibinfo{journal}{Nat. Photonics} \textbf{\bibinfo{volume}{5}},
  \bibinfo{pages}{222} (\bibinfo{year}{2011}).

\bibitem[{\citenamefont{Fr{\"o}wis and D{\"u}r}(2011)}]{frowis2011stable}
\bibinfo{author}{\bibfnamefont{F.}~\bibnamefont{Fr{\"o}wis}} \bibnamefont{and}
  \bibinfo{author}{\bibfnamefont{W.}~\bibnamefont{D{\"u}r}},
  \bibinfo{journal}{Phys. Rev. Lett.} \textbf{\bibinfo{volume}{106}},
  \bibinfo{pages}{110402} (\bibinfo{year}{2011}).

\bibitem[{\citenamefont{Demkowicz-Dobrza{\'n}ski
  et~al.}(2012)\citenamefont{Demkowicz-Dobrza{\'n}ski, Ko{\l}ody{\'n}ski, and
  Gu{\c{t}}{\u{a}}}}]{demkowicz2012elusive}
\bibinfo{author}{\bibfnamefont{R.}~\bibnamefont{Demkowicz-Dobrza{\'n}ski}},
  \bibinfo{author}{\bibfnamefont{J.}~\bibnamefont{Ko{\l}ody{\'n}ski}},
  \bibnamefont{and}
  \bibinfo{author}{\bibfnamefont{M.}~\bibnamefont{Gu{\c{t}}{\u{a}}}},
  \bibinfo{journal}{Nat. Commun.} \textbf{\bibinfo{volume}{3}},
  \bibinfo{pages}{1} (\bibinfo{year}{2012}).

\bibitem[{\citenamefont{Wang et~al.}(2018)\citenamefont{Wang, Wang, Zhan, Bian,
  Li, Sanders, and Xue}}]{wang2018entanglement}
\bibinfo{author}{\bibfnamefont{K.}~\bibnamefont{Wang}},
  \bibinfo{author}{\bibfnamefont{X.}~\bibnamefont{Wang}},
  \bibinfo{author}{\bibfnamefont{X.}~\bibnamefont{Zhan}},
  \bibinfo{author}{\bibfnamefont{Z.}~\bibnamefont{Bian}},
  \bibinfo{author}{\bibfnamefont{J.}~\bibnamefont{Li}},
  \bibinfo{author}{\bibfnamefont{B.~C.} \bibnamefont{Sanders}},
  \bibnamefont{and} \bibinfo{author}{\bibfnamefont{P.}~\bibnamefont{Xue}},
  \bibinfo{journal}{Phys. Rev. A} \textbf{\bibinfo{volume}{97}},
  \bibinfo{pages}{042112} (\bibinfo{year}{2018}).

\bibitem[{\citenamefont{Kwon et~al.}(2019)\citenamefont{Kwon, Tan, Volkoff, and
  Jeong}}]{kwon2019nonclassicality}
\bibinfo{author}{\bibfnamefont{H.}~\bibnamefont{Kwon}},
  \bibinfo{author}{\bibfnamefont{K.~C.} \bibnamefont{Tan}},
  \bibinfo{author}{\bibfnamefont{T.}~\bibnamefont{Volkoff}}, \bibnamefont{and}
  \bibinfo{author}{\bibfnamefont{H.}~\bibnamefont{Jeong}},
  \bibinfo{journal}{Phys. Rev. Lett.} \textbf{\bibinfo{volume}{122}},
  \bibinfo{pages}{040503} (\bibinfo{year}{2019}).

\bibitem[{\citenamefont{Raghunandan et~al.}(2018)\citenamefont{Raghunandan,
  Wrachtrup, and Weimer}}]{raghunandan2018high}
\bibinfo{author}{\bibfnamefont{M.}~\bibnamefont{Raghunandan}},
  \bibinfo{author}{\bibfnamefont{J.}~\bibnamefont{Wrachtrup}},
  \bibnamefont{and} \bibinfo{author}{\bibfnamefont{H.}~\bibnamefont{Weimer}},
  \bibinfo{journal}{Phys. Rev. Lett.} \textbf{\bibinfo{volume}{120}},
  \bibinfo{pages}{150501} (\bibinfo{year}{2018}).

\bibitem[{\citenamefont{Heugel et~al.}(2019)\citenamefont{Heugel, Biondi,
  Zilberberg, and Chitra}}]{heugel2019quantum}
\bibinfo{author}{\bibfnamefont{T.~L.} \bibnamefont{Heugel}},
  \bibinfo{author}{\bibfnamefont{M.}~\bibnamefont{Biondi}},
  \bibinfo{author}{\bibfnamefont{O.}~\bibnamefont{Zilberberg}},
  \bibnamefont{and} \bibinfo{author}{\bibfnamefont{R.}~\bibnamefont{Chitra}},
  \bibinfo{journal}{Phys. Rev. Lett.} \textbf{\bibinfo{volume}{123}},
  \bibinfo{pages}{173601} (\bibinfo{year}{2019}).

\bibitem[{\citenamefont{Yang and Jacob}(2019)}]{yang2019engineering}
\bibinfo{author}{\bibfnamefont{L.-P.} \bibnamefont{Yang}} \bibnamefont{and}
  \bibinfo{author}{\bibfnamefont{Z.}~\bibnamefont{Jacob}},
  \bibinfo{journal}{J. Appl. Phys.} \textbf{\bibinfo{volume}{126}},
  \bibinfo{pages}{174502} (\bibinfo{year}{2019}).

\bibitem[{\citenamefont{Zanardi and Paunkovi{\'c}}(2006)}]{zanardi2006ground}
\bibinfo{author}{\bibfnamefont{P.}~\bibnamefont{Zanardi}} \bibnamefont{and}
  \bibinfo{author}{\bibfnamefont{N.}~\bibnamefont{Paunkovi{\'c}}},
  \bibinfo{journal}{Phys. Rev. E} \textbf{\bibinfo{volume}{74}},
  \bibinfo{pages}{031123} (\bibinfo{year}{2006}).

\bibitem[{\citenamefont{Zanardi et~al.}(2007)\citenamefont{Zanardi, Quan, Wang,
  and Sun}}]{zanardi2007mixed}
\bibinfo{author}{\bibfnamefont{P.}~\bibnamefont{Zanardi}},
  \bibinfo{author}{\bibfnamefont{H.}~\bibnamefont{Quan}},
  \bibinfo{author}{\bibfnamefont{X.}~\bibnamefont{Wang}}, \bibnamefont{and}
  \bibinfo{author}{\bibfnamefont{C.}~\bibnamefont{Sun}},
  \bibinfo{journal}{Phys. Rev. A} \textbf{\bibinfo{volume}{75}},
  \bibinfo{pages}{032109} (\bibinfo{year}{2007}).

\bibitem[{\citenamefont{Gu et~al.}(2008)\citenamefont{Gu, Kwok, Ning, Lin
  et~al.}}]{gu2008fidelity}
\bibinfo{author}{\bibfnamefont{S.-J.} \bibnamefont{Gu}},
  \bibinfo{author}{\bibfnamefont{H.-M.} \bibnamefont{Kwok}},
  \bibinfo{author}{\bibfnamefont{W.-Q.} \bibnamefont{Ning}},
  \bibinfo{author}{\bibfnamefont{H.-Q.} \bibnamefont{Lin}},
  \bibnamefont{et~al.}, \bibinfo{journal}{Phys. Rev. B}
  \textbf{\bibinfo{volume}{77}}, \bibinfo{pages}{245109}
  (\bibinfo{year}{2008}).

\bibitem[{\citenamefont{Zanardi et~al.}(2008)\citenamefont{Zanardi, Paris, and
  Venuti}}]{zanardi2008quantum}
\bibinfo{author}{\bibfnamefont{P.}~\bibnamefont{Zanardi}},
  \bibinfo{author}{\bibfnamefont{M.~G.} \bibnamefont{Paris}}, \bibnamefont{and}
  \bibinfo{author}{\bibfnamefont{L.~C.} \bibnamefont{Venuti}},
  \bibinfo{journal}{Phys. Rev. A} \textbf{\bibinfo{volume}{78}},
  \bibinfo{pages}{042105} (\bibinfo{year}{2008}).

\bibitem[{\citenamefont{Invernizzi et~al.}(2008)\citenamefont{Invernizzi,
  Korbman, Venuti, and Paris}}]{invernizzi2008optimal}
\bibinfo{author}{\bibfnamefont{C.}~\bibnamefont{Invernizzi}},
  \bibinfo{author}{\bibfnamefont{M.}~\bibnamefont{Korbman}},
  \bibinfo{author}{\bibfnamefont{L.~C.} \bibnamefont{Venuti}},
  \bibnamefont{and} \bibinfo{author}{\bibfnamefont{M.~G.} \bibnamefont{Paris}},
  \bibinfo{journal}{Phys. Rev. A} \textbf{\bibinfo{volume}{78}},
  \bibinfo{pages}{042106} (\bibinfo{year}{2008}).

\bibitem[{\citenamefont{Gu}(2010)}]{gu2010fidelity}
\bibinfo{author}{\bibfnamefont{S.-J.} \bibnamefont{Gu}}, \bibinfo{journal}{Int.
  J. Mod. Phys. B} \textbf{\bibinfo{volume}{24}}, \bibinfo{pages}{4371}
  (\bibinfo{year}{2010}).

\bibitem[{\citenamefont{Gammelmark and M{\o}lmer}(2011)}]{gammelmark2011phase}
\bibinfo{author}{\bibfnamefont{S.}~\bibnamefont{Gammelmark}} \bibnamefont{and}
  \bibinfo{author}{\bibfnamefont{K.}~\bibnamefont{M{\o}lmer}},
  \bibinfo{journal}{New J. Phys.} \textbf{\bibinfo{volume}{13}},
  \bibinfo{pages}{053035} (\bibinfo{year}{2011}).

\bibitem[{\citenamefont{Skotiniotis et~al.}(2015)\citenamefont{Skotiniotis,
  Sekatski, and D{\"u}r}}]{skotiniotis2015quantum}
\bibinfo{author}{\bibfnamefont{M.}~\bibnamefont{Skotiniotis}},
  \bibinfo{author}{\bibfnamefont{P.}~\bibnamefont{Sekatski}}, \bibnamefont{and}
  \bibinfo{author}{\bibfnamefont{W.}~\bibnamefont{D{\"u}r}},
  \bibinfo{journal}{New J. Phys.} \textbf{\bibinfo{volume}{17}},
  \bibinfo{pages}{073032} (\bibinfo{year}{2015}).

\bibitem[{\citenamefont{Rams et~al.}(2018)\citenamefont{Rams, Sierant, Dutta,
  Horodecki, and Zakrzewski}}]{rams2018limits}
\bibinfo{author}{\bibfnamefont{M.~M.} \bibnamefont{Rams}},
  \bibinfo{author}{\bibfnamefont{P.}~\bibnamefont{Sierant}},
  \bibinfo{author}{\bibfnamefont{O.}~\bibnamefont{Dutta}},
  \bibinfo{author}{\bibfnamefont{P.}~\bibnamefont{Horodecki}},
  \bibnamefont{and}
  \bibinfo{author}{\bibfnamefont{J.}~\bibnamefont{Zakrzewski}},
  \bibinfo{journal}{Phys. Rev. X} \textbf{\bibinfo{volume}{8}},
  \bibinfo{pages}{021022} (\bibinfo{year}{2018}).

\bibitem[{\citenamefont{Wei}(2019)}]{wei2019fidelity}
\bibinfo{author}{\bibfnamefont{B.-B.} \bibnamefont{Wei}},
  \bibinfo{journal}{Phys. Rev. A} \textbf{\bibinfo{volume}{99}},
  \bibinfo{pages}{042117} (\bibinfo{year}{2019}).

\bibitem[{\citenamefont{Chu et~al.}(2021)\citenamefont{Chu, Zhang, Yu, and
  Cai}}]{chu2021dynamic}
\bibinfo{author}{\bibfnamefont{Y.}~\bibnamefont{Chu}},
  \bibinfo{author}{\bibfnamefont{S.}~\bibnamefont{Zhang}},
  \bibinfo{author}{\bibfnamefont{B.}~\bibnamefont{Yu}}, \bibnamefont{and}
  \bibinfo{author}{\bibfnamefont{J.}~\bibnamefont{Cai}},
  \bibinfo{journal}{Phys. Rev. Lett.} \textbf{\bibinfo{volume}{126}},
  \bibinfo{pages}{010502} (\bibinfo{year}{2021}).

\bibitem[{\citenamefont{Liu et~al.}(2021)\citenamefont{Liu, Chen, Jiang, Yang,
  Wu, Li, Yuan, Peng, and Du}}]{liu2021experimental}
\bibinfo{author}{\bibfnamefont{R.}~\bibnamefont{Liu}},
  \bibinfo{author}{\bibfnamefont{Y.}~\bibnamefont{Chen}},
  \bibinfo{author}{\bibfnamefont{M.}~\bibnamefont{Jiang}},
  \bibinfo{author}{\bibfnamefont{X.}~\bibnamefont{Yang}},
  \bibinfo{author}{\bibfnamefont{Z.}~\bibnamefont{Wu}},
  \bibinfo{author}{\bibfnamefont{Y.}~\bibnamefont{Li}},
  \bibinfo{author}{\bibfnamefont{H.}~\bibnamefont{Yuan}},
  \bibinfo{author}{\bibfnamefont{X.}~\bibnamefont{Peng}}, \bibnamefont{and}
  \bibinfo{author}{\bibfnamefont{J.}~\bibnamefont{Du}}, \bibinfo{journal}{npj
  Quantum Information} \textbf{\bibinfo{volume}{7}}, \bibinfo{pages}{170}
  (\bibinfo{year}{2021}).

\bibitem[{\citenamefont{Montenegro et~al.}(2021)\citenamefont{Montenegro,
  Mishra, and Bayat}}]{montenegro2021global}
\bibinfo{author}{\bibfnamefont{V.}~\bibnamefont{Montenegro}},
  \bibinfo{author}{\bibfnamefont{U.}~\bibnamefont{Mishra}}, \bibnamefont{and}
  \bibinfo{author}{\bibfnamefont{A.}~\bibnamefont{Bayat}},
  \bibinfo{journal}{Phys. Rev. Lett.} \textbf{\bibinfo{volume}{126}},
  \bibinfo{pages}{200501} (\bibinfo{year}{2021}).

\bibitem[{\citenamefont{Mirkhalaf et~al.}(2021)\citenamefont{Mirkhalaf, Orenes,
  Mitchell, and Witkowska}}]{mirkhalaf2021criticality}
\bibinfo{author}{\bibfnamefont{S.~S.} \bibnamefont{Mirkhalaf}},
  \bibinfo{author}{\bibfnamefont{D.~B.} \bibnamefont{Orenes}},
  \bibinfo{author}{\bibfnamefont{M.~W.} \bibnamefont{Mitchell}},
  \bibnamefont{and}
  \bibinfo{author}{\bibfnamefont{E.}~\bibnamefont{Witkowska}},
  \bibinfo{journal}{Phys. Rev. A} \textbf{\bibinfo{volume}{103}},
  \bibinfo{pages}{023317} (\bibinfo{year}{2021}).

\bibitem[{\citenamefont{Di~Candia et~al.}(2021)\citenamefont{Di~Candia,
  Minganti, Petrovnin, Paraoanu, and Felicetti}}]{di2021critical}
\bibinfo{author}{\bibfnamefont{R.}~\bibnamefont{Di~Candia}},
  \bibinfo{author}{\bibfnamefont{F.}~\bibnamefont{Minganti}},
  \bibinfo{author}{\bibfnamefont{K.}~\bibnamefont{Petrovnin}},
  \bibinfo{author}{\bibfnamefont{G.}~\bibnamefont{Paraoanu}}, \bibnamefont{and}
  \bibinfo{author}{\bibfnamefont{S.}~\bibnamefont{Felicetti}},
  \bibinfo{journal}{npj
  	Quantum Information} \textbf{\bibinfo{volume}{9}}, \bibinfo{pages}{23}
  (\bibinfo{year}{2023})

\bibitem[{\citenamefont{Baumann et~al.}(2010)\citenamefont{Baumann, Guerlin,
  Brennecke, and Esslinger}}]{baumann2010dicke}
\bibinfo{author}{\bibfnamefont{K.}~\bibnamefont{Baumann}},
  \bibinfo{author}{\bibfnamefont{C.}~\bibnamefont{Guerlin}},
  \bibinfo{author}{\bibfnamefont{F.}~\bibnamefont{Brennecke}},
  \bibnamefont{and}
  \bibinfo{author}{\bibfnamefont{T.}~\bibnamefont{Esslinger}},
  \bibinfo{journal}{Nature (London)} \textbf{\bibinfo{volume}{464}},
  \bibinfo{pages}{1301} (\bibinfo{year}{2010}).

\bibitem[{\citenamefont{Baden et~al.}(2014)\citenamefont{Baden, Arnold,
  Grimsmo, Parkins, and Barrett}}]{baden2014realization}
\bibinfo{author}{\bibfnamefont{M.~P.} \bibnamefont{Baden}},
  \bibinfo{author}{\bibfnamefont{K.~J.} \bibnamefont{Arnold}},
  \bibinfo{author}{\bibfnamefont{A.~L.} \bibnamefont{Grimsmo}},
  \bibinfo{author}{\bibfnamefont{S.}~\bibnamefont{Parkins}}, \bibnamefont{and}
  \bibinfo{author}{\bibfnamefont{M.~D.} \bibnamefont{Barrett}},
  \bibinfo{journal}{Phys. Rev. Lett.} \textbf{\bibinfo{volume}{113}},
  \bibinfo{pages}{020408} (\bibinfo{year}{2014}).

\bibitem[{\citenamefont{Klinder et~al.}(2015)\citenamefont{Klinder, Ke{\ss}ler,
  Wolke, Mathey, and Hemmerich}}]{klinder2015dynamical}
\bibinfo{author}{\bibfnamefont{J.}~\bibnamefont{Klinder}},
  \bibinfo{author}{\bibfnamefont{H.}~\bibnamefont{Ke{\ss}ler}},
  \bibinfo{author}{\bibfnamefont{M.}~\bibnamefont{Wolke}},
  \bibinfo{author}{\bibfnamefont{L.}~\bibnamefont{Mathey}}, \bibnamefont{and}
  \bibinfo{author}{\bibfnamefont{A.}~\bibnamefont{Hemmerich}},
  \bibinfo{journal}{Proc. Natl. Acad. Sci. U.S.A.}
  \textbf{\bibinfo{volume}{112}}, \bibinfo{pages}{3290} (\bibinfo{year}{2015}).

\bibitem[{\citenamefont{Rodriguez et~al.}(2017)\citenamefont{Rodriguez,
  Casteels, Storme, Zambon, Sagnes, Le~Gratiet, Galopin, Lema{\^\i}tre, Amo,
  Ciuti et~al.}}]{rodriguez2017probing}
\bibinfo{author}{\bibfnamefont{S.}~\bibnamefont{Rodriguez}},
  \bibinfo{author}{\bibfnamefont{W.}~\bibnamefont{Casteels}},
  \bibinfo{author}{\bibfnamefont{F.}~\bibnamefont{Storme}},
  \bibinfo{author}{\bibfnamefont{N.~C.} \bibnamefont{Zambon}},
  \bibinfo{author}{\bibfnamefont{I.}~\bibnamefont{Sagnes}},
  \bibinfo{author}{\bibfnamefont{L.}~\bibnamefont{Le~Gratiet}},
  \bibinfo{author}{\bibfnamefont{E.}~\bibnamefont{Galopin}},
  \bibinfo{author}{\bibfnamefont{A.}~\bibnamefont{Lema{\^\i}tre}},
  \bibinfo{author}{\bibfnamefont{A.}~\bibnamefont{Amo}},
  \bibinfo{author}{\bibfnamefont{C.}~\bibnamefont{Ciuti}},
  \bibnamefont{et~al.}, \bibinfo{journal}{Phys. Rev. Lett.}
  \textbf{\bibinfo{volume}{118}}, \bibinfo{pages}{247402}
  (\bibinfo{year}{2017}).

\bibitem[{\citenamefont{Fitzpatrick et~al.}(2017)\citenamefont{Fitzpatrick,
  Sundaresan, Li, Koch, and Houck}}]{fitzpatrick2017observation}
\bibinfo{author}{\bibfnamefont{M.}~\bibnamefont{Fitzpatrick}},
  \bibinfo{author}{\bibfnamefont{N.~M.} \bibnamefont{Sundaresan}},
  \bibinfo{author}{\bibfnamefont{A.~C.} \bibnamefont{Li}},
  \bibinfo{author}{\bibfnamefont{J.}~\bibnamefont{Koch}}, \bibnamefont{and}
  \bibinfo{author}{\bibfnamefont{A.~A.} \bibnamefont{Houck}},
  \bibinfo{journal}{Phys. Rev. X} \textbf{\bibinfo{volume}{7}},
  \bibinfo{pages}{011016} (\bibinfo{year}{2017}).

\bibitem[{\citenamefont{Fink et~al.}(2017)\citenamefont{Fink, Dombi, Vukics,
  Wallraff, and Domokos}}]{fink2017observation}
\bibinfo{author}{\bibfnamefont{J.~M.} \bibnamefont{Fink}},
  \bibinfo{author}{\bibfnamefont{A.}~\bibnamefont{Dombi}},
  \bibinfo{author}{\bibfnamefont{A.}~\bibnamefont{Vukics}},
  \bibinfo{author}{\bibfnamefont{A.}~\bibnamefont{Wallraff}}, \bibnamefont{and}
  \bibinfo{author}{\bibfnamefont{P.}~\bibnamefont{Domokos}},
  \bibinfo{journal}{Phys. Rev. X} \textbf{\bibinfo{volume}{7}},
  \bibinfo{pages}{011012} (\bibinfo{year}{2017}).

\bibitem[{\citenamefont{Ilias et~al.}(2022)\citenamefont{Ilias, Yang, Huelga,
  and Plenio}}]{ilias2022criticality}
\bibinfo{author}{\bibfnamefont{T.}~\bibnamefont{Ilias}},
  \bibinfo{author}{\bibfnamefont{D.}~\bibnamefont{Yang}},
  \bibinfo{author}{\bibfnamefont{S.~F.} \bibnamefont{Huelga}},
  \bibnamefont{and} \bibinfo{author}{\bibfnamefont{M.~B.}
  \bibnamefont{Plenio}}, \bibinfo{journal}{PRX Quantum}
  \textbf{\bibinfo{volume}{3}}, \bibinfo{pages}{010354} (\bibinfo{year}{2022}).

\bibitem[{\citenamefont{Montenegro et~al.}(2023)\citenamefont{Montenegro,
  Genoni, Bayat, and Paris}}]{montenegro2023quantum}
\bibinfo{author}{\bibfnamefont{V.}~\bibnamefont{Montenegro}},
  \bibinfo{author}{\bibfnamefont{M.}~\bibnamefont{Genoni}},
  \bibinfo{author}{\bibfnamefont{A.}~\bibnamefont{Bayat}}, \bibnamefont{and}
  \bibinfo{author}{\bibfnamefont{M.}~\bibnamefont{Paris}},
  \bibinfo{journal}{arXiv:2301.02103}.

\bibitem[{\citenamefont{Budich and Bergholtz}(2020)}]{budich2020non}
\bibinfo{author}{\bibfnamefont{J.~C.} \bibnamefont{Budich}} \bibnamefont{and}
  \bibinfo{author}{\bibfnamefont{E.~J.} \bibnamefont{Bergholtz}},
  \bibinfo{journal}{Phys. Rev. Lett.} \textbf{\bibinfo{volume}{125}},
  \bibinfo{pages}{180403} (\bibinfo{year}{2020}).

\bibitem[{\citenamefont{Sarkar et~al.}(2022)\citenamefont{Sarkar, Mukhopadhyay,
  Alase, and Bayat}}]{sarkar2022free}
\bibinfo{author}{\bibfnamefont{S.}~\bibnamefont{Sarkar}},
  \bibinfo{author}{\bibfnamefont{C.}~\bibnamefont{Mukhopadhyay}},
  \bibinfo{author}{\bibfnamefont{A.}~\bibnamefont{Alase}}, \bibnamefont{and}
  \bibinfo{author}{\bibfnamefont{A.}~\bibnamefont{Bayat}},
  \bibinfo{journal}{Phys. Rev. Lett.} \textbf{\bibinfo{volume}{129}},
  \bibinfo{pages}{090503} (\bibinfo{year}{2022}).

\bibitem[{\citenamefont{Koch and Budich}(2022)}]{koch2022quantum}
\bibinfo{author}{\bibfnamefont{F.}~\bibnamefont{Koch}} \bibnamefont{and}
  \bibinfo{author}{\bibfnamefont{J.~C.} \bibnamefont{Budich}},
  \bibinfo{journal}{Phys. Rev. Res.} \textbf{\bibinfo{volume}{4}},
  \bibinfo{pages}{013113} (\bibinfo{year}{2022}).

\bibitem[{\citenamefont{Yu et~al.}(2022)\citenamefont{Yu, Li, Chu, Mera,
  {\"U}nal, Yang, Liu, Goldman, and Cai}}]{yu2022experimental}
\bibinfo{author}{\bibfnamefont{M.}~\bibnamefont{Yu}},
  \bibinfo{author}{\bibfnamefont{X.}~\bibnamefont{Li}},
  \bibinfo{author}{\bibfnamefont{Y.}~\bibnamefont{Chu}},
  \bibinfo{author}{\bibfnamefont{B.}~\bibnamefont{Mera}},
  \bibinfo{author}{\bibfnamefont{F.~N.} \bibnamefont{{\"U}nal}},
  \bibinfo{author}{\bibfnamefont{P.}~\bibnamefont{Yang}},
  \bibinfo{author}{\bibfnamefont{Y.}~\bibnamefont{Liu}},
  \bibinfo{author}{\bibfnamefont{N.}~\bibnamefont{Goldman}}, \bibnamefont{and}
  \bibinfo{author}{\bibfnamefont{J.}~\bibnamefont{Cai}},
  \bibinfo{journal}{arXiv:2206.00546}.

\bibitem[{\citenamefont{Dooley}(2021)}]{dooley2021robust}
\bibinfo{author}{\bibfnamefont{S.}~\bibnamefont{Dooley}}, \bibinfo{journal}{PRX
  Quantum} \textbf{\bibinfo{volume}{2}}, \bibinfo{pages}{020330}
  (\bibinfo{year}{2021}).

\bibitem[{\citenamefont{Desaules et~al.}(2021)\citenamefont{Desaules, Hudomal,
  Turner, and Papi{\'c}}}]{desaules2021proposal}
\bibinfo{author}{\bibfnamefont{J.-Y.} \bibnamefont{Desaules}},
  \bibinfo{author}{\bibfnamefont{A.}~\bibnamefont{Hudomal}},
  \bibinfo{author}{\bibfnamefont{C.~J.} \bibnamefont{Turner}},
  \bibnamefont{and}
  \bibinfo{author}{\bibfnamefont{Z.}~\bibnamefont{Papi{\'c}}},
  \bibinfo{journal}{Phys. Rev. Lett.} \textbf{\bibinfo{volume}{126}},
  \bibinfo{pages}{210601} (\bibinfo{year}{2021}).

\bibitem[{\citenamefont{Yoshinaga et~al.}(2022)\citenamefont{Yoshinaga,
  Matsuzaki, and Hamazaki}}]{yoshinaga2022quantum}
\bibinfo{author}{\bibfnamefont{A.}~\bibnamefont{Yoshinaga}},
  \bibinfo{author}{\bibfnamefont{Y.}~\bibnamefont{Matsuzaki}},
  \bibnamefont{and} \bibinfo{author}{\bibfnamefont{R.}~\bibnamefont{Hamazaki}},
  \bibinfo{journal}{arXiv:2211.09567}.

\bibitem[{\citenamefont{Dooley et~al.}(2023)\citenamefont{Dooley, Pappalardi,
  and Goold}}]{dooley2023entanglement}
\bibinfo{author}{\bibfnamefont{S.}~\bibnamefont{Dooley}},
  \bibinfo{author}{\bibfnamefont{S.}~\bibnamefont{Pappalardi}},
  \bibnamefont{and} \bibinfo{author}{\bibfnamefont{J.}~\bibnamefont{Goold}},
  \bibinfo{journal}{Phys. Rev. B} \textbf{\bibinfo{volume}{107}},
  \bibinfo{pages}{035123} (\bibinfo{year}{2023}).

\bibitem[{\citenamefont{Mishra and Bayat}(2021)}]{mishra2021driving}
\bibinfo{author}{\bibfnamefont{U.}~\bibnamefont{Mishra}} \bibnamefont{and}
  \bibinfo{author}{\bibfnamefont{A.}~\bibnamefont{Bayat}},
  \bibinfo{journal}{Phys. Rev. Lett.} \textbf{\bibinfo{volume}{127}},
  \bibinfo{pages}{080504} (\bibinfo{year}{2021}).

\bibitem[{\citenamefont{Mishra and Bayat}(2022)}]{mishra2022integrable}
\bibinfo{author}{\bibfnamefont{U.}~\bibnamefont{Mishra}} \bibnamefont{and}
  \bibinfo{author}{\bibfnamefont{A.}~\bibnamefont{Bayat}},
  \bibinfo{journal}{Sci. Rep.} \textbf{\bibinfo{volume}{12}},
  \bibinfo{pages}{14760} (\bibinfo{year}{2022}).

\bibitem[{\citenamefont{Wiseman}(1995)}]{wiseman1995adaptive}
\bibinfo{author}{\bibfnamefont{H.~M.} \bibnamefont{Wiseman}},
  \bibinfo{journal}{Phys. Rev. Lett.} \textbf{\bibinfo{volume}{75}},
  \bibinfo{pages}{4587} (\bibinfo{year}{1995}).

\bibitem[{\citenamefont{Armen et~al.}(2002)\citenamefont{Armen, Au, Stockton,
  Doherty, and Mabuchi}}]{armen2002adaptive}
\bibinfo{author}{\bibfnamefont{M.~A.} \bibnamefont{Armen}},
  \bibinfo{author}{\bibfnamefont{J.~K.} \bibnamefont{Au}},
  \bibinfo{author}{\bibfnamefont{J.~K.} \bibnamefont{Stockton}},
  \bibinfo{author}{\bibfnamefont{A.~C.} \bibnamefont{Doherty}},
  \bibnamefont{and} \bibinfo{author}{\bibfnamefont{H.}~\bibnamefont{Mabuchi}},
  \bibinfo{journal}{Phys. Rev. Lett.} \textbf{\bibinfo{volume}{89}},
  \bibinfo{pages}{133602} (\bibinfo{year}{2002}).

\bibitem[{\citenamefont{Fujiwara}(2006)}]{fujiwara2006strong}
\bibinfo{author}{\bibfnamefont{A.}~\bibnamefont{Fujiwara}},
  \bibinfo{journal}{J. Phys. A Math. Gen.} \textbf{\bibinfo{volume}{39}},
  \bibinfo{pages}{12489} (\bibinfo{year}{2006}).

\bibitem[{\citenamefont{Higgins et~al.}(2007)\citenamefont{Higgins, Berry,
  Bartlett, Wiseman, and Pryde}}]{higgins2007entanglement}
\bibinfo{author}{\bibfnamefont{B.~L.} \bibnamefont{Higgins}},
  \bibinfo{author}{\bibfnamefont{D.~W.} \bibnamefont{Berry}},
  \bibinfo{author}{\bibfnamefont{S.~D.} \bibnamefont{Bartlett}},
  \bibinfo{author}{\bibfnamefont{H.~M.} \bibnamefont{Wiseman}},
  \bibnamefont{and} \bibinfo{author}{\bibfnamefont{G.~J.} \bibnamefont{Pryde}},
  \bibinfo{journal}{Nature (London)} \textbf{\bibinfo{volume}{450}},
  \bibinfo{pages}{393} (\bibinfo{year}{2007}).

\bibitem[{\citenamefont{Berry et~al.}(2009)\citenamefont{Berry, Higgins,
  Bartlett, Mitchell, Pryde, and Wiseman}}]{berry2009perform}
\bibinfo{author}{\bibfnamefont{D.~W.} \bibnamefont{Berry}},
  \bibinfo{author}{\bibfnamefont{B.~L.} \bibnamefont{Higgins}},
  \bibinfo{author}{\bibfnamefont{S.~D.} \bibnamefont{Bartlett}},
  \bibinfo{author}{\bibfnamefont{M.~W.} \bibnamefont{Mitchell}},
  \bibinfo{author}{\bibfnamefont{G.~J.} \bibnamefont{Pryde}}, \bibnamefont{and}
  \bibinfo{author}{\bibfnamefont{H.~M.} \bibnamefont{Wiseman}},
  \bibinfo{journal}{Phys. Rev. A} \textbf{\bibinfo{volume}{80}},
  \bibinfo{pages}{052114} (\bibinfo{year}{2009}).

\bibitem[{\citenamefont{Said et~al.}(2011)\citenamefont{Said, Berry, and
  Twamley}}]{said2011nanoscale}
\bibinfo{author}{\bibfnamefont{R.}~\bibnamefont{Said}},
  \bibinfo{author}{\bibfnamefont{D.}~\bibnamefont{Berry}}, \bibnamefont{and}
  \bibinfo{author}{\bibfnamefont{J.}~\bibnamefont{Twamley}},
  \bibinfo{journal}{Phys. Rev. B} \textbf{\bibinfo{volume}{83}},
  \bibinfo{pages}{125410} (\bibinfo{year}{2011}).

\bibitem[{\citenamefont{Okamoto et~al.}(2012)\citenamefont{Okamoto, Iefuji,
  Oyama, Yamagata, Imai, Fujiwara, and Takeuchi}}]{okamoto2012experimental}
\bibinfo{author}{\bibfnamefont{R.}~\bibnamefont{Okamoto}},
  \bibinfo{author}{\bibfnamefont{M.}~\bibnamefont{Iefuji}},
  \bibinfo{author}{\bibfnamefont{S.}~\bibnamefont{Oyama}},
  \bibinfo{author}{\bibfnamefont{K.}~\bibnamefont{Yamagata}},
  \bibinfo{author}{\bibfnamefont{H.}~\bibnamefont{Imai}},
  \bibinfo{author}{\bibfnamefont{A.}~\bibnamefont{Fujiwara}}, \bibnamefont{and}
  \bibinfo{author}{\bibfnamefont{S.}~\bibnamefont{Takeuchi}},
  \bibinfo{journal}{Phys. Rev. Lett.} \textbf{\bibinfo{volume}{109}},
  \bibinfo{pages}{130404} (\bibinfo{year}{2012}).

\bibitem[{\citenamefont{Bonato et~al.}(2016)\citenamefont{Bonato, Blok, Dinani,
  Berry, Markham, Twitchen, and Hanson}}]{bonato2016optimized}
\bibinfo{author}{\bibfnamefont{C.}~\bibnamefont{Bonato}},
  \bibinfo{author}{\bibfnamefont{M.~S.} \bibnamefont{Blok}},
  \bibinfo{author}{\bibfnamefont{H.~T.} \bibnamefont{Dinani}},
  \bibinfo{author}{\bibfnamefont{D.~W.} \bibnamefont{Berry}},
  \bibinfo{author}{\bibfnamefont{M.~L.} \bibnamefont{Markham}},
  \bibinfo{author}{\bibfnamefont{D.~J.} \bibnamefont{Twitchen}},
  \bibnamefont{and} \bibinfo{author}{\bibfnamefont{R.}~\bibnamefont{Hanson}},
  \bibinfo{journal}{Nat. Nanotechnol.} \textbf{\bibinfo{volume}{11}},
  \bibinfo{pages}{247} (\bibinfo{year}{2016}).

\bibitem[{\citenamefont{Okamoto et~al.}(2017)\citenamefont{Okamoto, Oyama,
  Yamagata, Fujiwara, and Takeuchi}}]{okamoto2017experimental}
\bibinfo{author}{\bibfnamefont{R.}~\bibnamefont{Okamoto}},
  \bibinfo{author}{\bibfnamefont{S.}~\bibnamefont{Oyama}},
  \bibinfo{author}{\bibfnamefont{K.}~\bibnamefont{Yamagata}},
  \bibinfo{author}{\bibfnamefont{A.}~\bibnamefont{Fujiwara}}, \bibnamefont{and}
  \bibinfo{author}{\bibfnamefont{S.}~\bibnamefont{Takeuchi}},
  \bibinfo{journal}{Phys. Rev. A} \textbf{\bibinfo{volume}{96}},
  \bibinfo{pages}{022124} (\bibinfo{year}{2017}).

\bibitem[{\citenamefont{Fern{\'a}ndez-Lorenzo and
  Porras}(2017)}]{fernandez2017quantum}
\bibinfo{author}{\bibfnamefont{S.}~\bibnamefont{Fern{\'a}ndez-Lorenzo}}
  \bibnamefont{and} \bibinfo{author}{\bibfnamefont{D.}~\bibnamefont{Porras}},
  \bibinfo{journal}{Phys. Rev. A} \textbf{\bibinfo{volume}{96}},
  \bibinfo{pages}{013817} (\bibinfo{year}{2017}).

\bibitem[{\citenamefont{Gammelmark and M{\o}lmer}(2014)}]{gammelmark2014fisher}
\bibinfo{author}{\bibfnamefont{S.}~\bibnamefont{Gammelmark}} \bibnamefont{and}
  \bibinfo{author}{\bibfnamefont{K.}~\bibnamefont{M{\o}lmer}},
  \bibinfo{journal}{Phys. Rev. Lett.} \textbf{\bibinfo{volume}{112}},
  \bibinfo{pages}{170401} (\bibinfo{year}{2014}).

\bibitem[{\citenamefont{Albarelli et~al.}(2017)\citenamefont{Albarelli, Rossi,
  Paris, and Genoni}}]{albarelli2017ultimate}
\bibinfo{author}{\bibfnamefont{F.}~\bibnamefont{Albarelli}},
  \bibinfo{author}{\bibfnamefont{M.~A.} \bibnamefont{Rossi}},
  \bibinfo{author}{\bibfnamefont{M.~G.} \bibnamefont{Paris}}, \bibnamefont{and}
  \bibinfo{author}{\bibfnamefont{M.~G.} \bibnamefont{Genoni}},
  \bibinfo{journal}{New J. Phys.} \textbf{\bibinfo{volume}{19}},
  \bibinfo{pages}{123011} (\bibinfo{year}{2017}).

\bibitem[{\citenamefont{Rossi et~al.}(2020)\citenamefont{Rossi, Albarelli,
  Tamascelli, and Genoni}}]{rossi2020noisy}
\bibinfo{author}{\bibfnamefont{M.~A.} \bibnamefont{Rossi}},
  \bibinfo{author}{\bibfnamefont{F.}~\bibnamefont{Albarelli}},
  \bibinfo{author}{\bibfnamefont{D.}~\bibnamefont{Tamascelli}},
  \bibnamefont{and} \bibinfo{author}{\bibfnamefont{M.~G.}
  \bibnamefont{Genoni}}, \bibinfo{journal}{Phys. Rev. Lett.}
  \textbf{\bibinfo{volume}{125}}, \bibinfo{pages}{200505}
  (\bibinfo{year}{2020}).

\bibitem[{\citenamefont{Yang et~al.}(2022)\citenamefont{Yang, Huelga, and
  Plenio}}]{yang2022efficient}
\bibinfo{author}{\bibfnamefont{D.}~\bibnamefont{Yang}},
  \bibinfo{author}{\bibfnamefont{S.~F.} \bibnamefont{Huelga}},
  \bibnamefont{and} \bibinfo{author}{\bibfnamefont{M.~B.}
  \bibnamefont{Plenio}}, \bibinfo{journal}{arXiv:2209.08777}.

\bibitem[{\citenamefont{Burgarth et~al.}(2015)\citenamefont{Burgarth,
  Giovannetti, Kato, and Yuasa}}]{burgarth2015quantum}
\bibinfo{author}{\bibfnamefont{D.}~\bibnamefont{Burgarth}},
  \bibinfo{author}{\bibfnamefont{V.}~\bibnamefont{Giovannetti}},
  \bibinfo{author}{\bibfnamefont{A.~N.} \bibnamefont{Kato}}, \bibnamefont{and}
  \bibinfo{author}{\bibfnamefont{K.}~\bibnamefont{Yuasa}},
  \bibinfo{journal}{New J. Phys.} \textbf{\bibinfo{volume}{17}},
  \bibinfo{pages}{113055} (\bibinfo{year}{2015}).

\bibitem[{\citenamefont{Montenegro et~al.}(2022)\citenamefont{Montenegro,
  Jones, Bose, and Bayat}}]{montenegro2022sequential}
\bibinfo{author}{\bibfnamefont{V.}~\bibnamefont{Montenegro}},
  \bibinfo{author}{\bibfnamefont{G.~S.} \bibnamefont{Jones}},
  \bibinfo{author}{\bibfnamefont{S.}~\bibnamefont{Bose}}, \bibnamefont{and}
  \bibinfo{author}{\bibfnamefont{A.}~\bibnamefont{Bayat}},
  \bibinfo{journal}{Phys. Rev. Lett.} \textbf{\bibinfo{volume}{129}},
  \bibinfo{pages}{120503} (\bibinfo{year}{2022}).

\bibitem[{\citenamefont{Boixo et~al.}(2007)\citenamefont{Boixo, Flammia, Caves,
  and Geremia}}]{boixo2007generalized}
\bibinfo{author}{\bibfnamefont{S.}~\bibnamefont{Boixo}},
  \bibinfo{author}{\bibfnamefont{S.~T.} \bibnamefont{Flammia}},
  \bibinfo{author}{\bibfnamefont{C.~M.} \bibnamefont{Caves}}, \bibnamefont{and}
  \bibinfo{author}{\bibfnamefont{J.~M.} \bibnamefont{Geremia}},
  \bibinfo{journal}{Phys. Rev. Lett.} \textbf{\bibinfo{volume}{98}},
  \bibinfo{pages}{090401} (\bibinfo{year}{2007}).

\bibitem[{\citenamefont{Roy and Braunstein}(2008)}]{roy2008exponentially}
\bibinfo{author}{\bibfnamefont{S.}~\bibnamefont{Roy}} \bibnamefont{and}
  \bibinfo{author}{\bibfnamefont{S.~L.} \bibnamefont{Braunstein}},
  \bibinfo{journal}{Phys. Rev. Lett.} \textbf{\bibinfo{volume}{100}},
  \bibinfo{pages}{220501} (\bibinfo{year}{2008}).

\bibitem[{\citenamefont{Beau and del Campo}(2017)}]{beau2017nonlinear}
\bibinfo{author}{\bibfnamefont{M.}~\bibnamefont{Beau}} \bibnamefont{and}
  \bibinfo{author}{\bibfnamefont{A.}~\bibnamefont{del Campo}},
  \bibinfo{journal}{Phys. Rev. Lett.} \textbf{\bibinfo{volume}{119}},
  \bibinfo{pages}{010403} (\bibinfo{year}{2017}).

\bibitem[{\citenamefont{Wannier}(1960)}]{wannier1960wave}
\bibinfo{author}{\bibfnamefont{G.~H.} \bibnamefont{Wannier}},
  \bibinfo{journal}{Phys. Rev.} \textbf{\bibinfo{volume}{117}},
  \bibinfo{pages}{432} (\bibinfo{year}{1960}).

\bibitem[{\citenamefont{Fukuyama et~al.}(1973)\citenamefont{Fukuyama, Bari, and
  Fogedby}}]{fukuyama1973tightly}
\bibinfo{author}{\bibfnamefont{H.}~\bibnamefont{Fukuyama}},
  \bibinfo{author}{\bibfnamefont{R.~A.} \bibnamefont{Bari}}, \bibnamefont{and}
  \bibinfo{author}{\bibfnamefont{H.~C.} \bibnamefont{Fogedby}},
  \bibinfo{journal}{Phys. Rev. B} \textbf{\bibinfo{volume}{8}},
  \bibinfo{pages}{5579} (\bibinfo{year}{1973}).

\bibitem[{\citenamefont{Holthaus et~al.}(1995)\citenamefont{Holthaus, Ristow,
  and Hone}}]{holthaus1995random}
\bibinfo{author}{\bibfnamefont{M.}~\bibnamefont{Holthaus}},
  \bibinfo{author}{\bibfnamefont{G.}~\bibnamefont{Ristow}}, \bibnamefont{and}
  \bibinfo{author}{\bibfnamefont{D.}~\bibnamefont{Hone}},
  \bibinfo{journal}{Europhys. Lett.} \textbf{\bibinfo{volume}{32}}, \bibinfo{pages}{241}
  (\bibinfo{year}{1995}).

\bibitem[{\citenamefont{Kolovsky and Korsch}(2003)}]{kolovsky2003bloch}
\bibinfo{author}{\bibfnamefont{A.~R.}~\bibnamefont{Kolovsky}} \bibnamefont{and}
  \bibinfo{author}{\bibfnamefont{H.}~\bibnamefont{Korsch}},
  \bibinfo{journal}{Phys. Rev. A} \textbf{\bibinfo{volume}{67}},
  \bibinfo{pages}{063601} (\bibinfo{year}{2003}).

\bibitem[{\citenamefont{Kolovsky}(2008)}]{kolovsky2008interplay}
\bibinfo{author}{\bibfnamefont{A.~R.} \bibnamefont{Kolovsky}},
  \bibinfo{journal}{Phys. Rev. Lett.} \textbf{\bibinfo{volume}{101}},
  \bibinfo{pages}{190602} (\bibinfo{year}{2008}).

\bibitem[{\citenamefont{Kolovsky and Bulgakov}(2013)}]{kolovsky2013wannier}
\bibinfo{author}{\bibfnamefont{A.~R.} \bibnamefont{Kolovsky}} \bibnamefont{and}
  \bibinfo{author}{\bibfnamefont{E.~N.} \bibnamefont{Bulgakov}},
  \bibinfo{journal}{Phys. Rev. A} \textbf{\bibinfo{volume}{87}},
  \bibinfo{pages}{033602} (\bibinfo{year}{2013}).

\bibitem[{\citenamefont{van Nieuwenburg et~al.}(2019)\citenamefont{van
  Nieuwenburg, Baum, and Refael}}]{van2019bloch}
\bibinfo{author}{\bibfnamefont{E.}~\bibnamefont{van Nieuwenburg}},
  \bibinfo{author}{\bibfnamefont{Y.}~\bibnamefont{Baum}}, \bibnamefont{and}
  \bibinfo{author}{\bibfnamefont{G.}~\bibnamefont{Refael}},
  \bibinfo{journal}{Proc. Natl. Acad. Sci. U.S.A.}
  \textbf{\bibinfo{volume}{116}}, \bibinfo{pages}{9269} (\bibinfo{year}{2019}).

\bibitem[{\citenamefont{Schulz et~al.}(2019)\citenamefont{Schulz, Hooley,
  Moessner, and Pollmann}}]{schulz2019stark}
\bibinfo{author}{\bibfnamefont{M.}~\bibnamefont{Schulz}},
  \bibinfo{author}{\bibfnamefont{C.}~\bibnamefont{Hooley}},
  \bibinfo{author}{\bibfnamefont{R.}~\bibnamefont{Moessner}}, \bibnamefont{and}
  \bibinfo{author}{\bibfnamefont{F.}~\bibnamefont{Pollmann}},
  \bibinfo{journal}{Phys. Rev. Lett.} \textbf{\bibinfo{volume}{122}},
  \bibinfo{pages}{040606} (\bibinfo{year}{2019}).

\bibitem[{\citenamefont{Wu and Eckardt}(2019)}]{wu2019bath}
\bibinfo{author}{\bibfnamefont{L.-N.} \bibnamefont{Wu}} \bibnamefont{and}
  \bibinfo{author}{\bibfnamefont{A.}~\bibnamefont{Eckardt}},
  \bibinfo{journal}{Phys. Rev. Lett.} \textbf{\bibinfo{volume}{123}},
  \bibinfo{pages}{030602} (\bibinfo{year}{2019}).

\bibitem[{\citenamefont{Bhakuni et~al.}(2020)\citenamefont{Bhakuni, Nehra, and
  Sharma}}]{bhakuni2020drive}
\bibinfo{author}{\bibfnamefont{D.~S.} \bibnamefont{Bhakuni}},
  \bibinfo{author}{\bibfnamefont{R.}~\bibnamefont{Nehra}}, \bibnamefont{and}
  \bibinfo{author}{\bibfnamefont{A.}~\bibnamefont{Sharma}},
  \bibinfo{journal}{Phys. Rev. B} \textbf{\bibinfo{volume}{102}},
  \bibinfo{pages}{024201} (\bibinfo{year}{2020}).

\bibitem[{\citenamefont{Bhakuni and Sharma}(2020)}]{bhakuni2020stability}
\bibinfo{author}{\bibfnamefont{D.~S.} \bibnamefont{Bhakuni}} \bibnamefont{and}
  \bibinfo{author}{\bibfnamefont{A.}~\bibnamefont{Sharma}},
  \bibinfo{journal}{Phys. Rev. B} \textbf{\bibinfo{volume}{102}},
  \bibinfo{pages}{085133} (\bibinfo{year}{2020}).

\bibitem[{\citenamefont{Yao and Zakrzewski}(2020)}]{yao2020many}
\bibinfo{author}{\bibfnamefont{R.}~\bibnamefont{Yao}} \bibnamefont{and}
  \bibinfo{author}{\bibfnamefont{J.}~\bibnamefont{Zakrzewski}},
  \bibinfo{journal}{Phys. Rev. B} \textbf{\bibinfo{volume}{102}},
  \bibinfo{pages}{104203} (\bibinfo{year}{2020}).

\bibitem[{\citenamefont{Chanda et~al.}(2020)\citenamefont{Chanda, Yao, and
  Zakrzewski}}]{chanda2020coexistence}
\bibinfo{author}{\bibfnamefont{T.}~\bibnamefont{Chanda}},
  \bibinfo{author}{\bibfnamefont{R.}~\bibnamefont{Yao}}, \bibnamefont{and}
  \bibinfo{author}{\bibfnamefont{J.}~\bibnamefont{Zakrzewski}},
  \bibinfo{journal}{Phys. Rev. Res.} \textbf{\bibinfo{volume}{2}},
  \bibinfo{pages}{032039(R)} (\bibinfo{year}{2020}).

\bibitem[{\citenamefont{Taylor et~al.}(2020)\citenamefont{Taylor, Schulz,
  Pollmann, and Moessner}}]{taylor2020experimental}
\bibinfo{author}{\bibfnamefont{S.~R.} \bibnamefont{Taylor}},
  \bibinfo{author}{\bibfnamefont{M.}~\bibnamefont{Schulz}},
  \bibinfo{author}{\bibfnamefont{F.}~\bibnamefont{Pollmann}}, \bibnamefont{and}
  \bibinfo{author}{\bibfnamefont{R.}~\bibnamefont{Moessner}},
  \bibinfo{journal}{Phys. Rev. B} \textbf{\bibinfo{volume}{102}},
  \bibinfo{pages}{054206} (\bibinfo{year}{2020}).

\bibitem[{\citenamefont{Wang et~al.}(2021)\citenamefont{Wang, Sun, and
  Fan}}]{wang2021stark}
\bibinfo{author}{\bibfnamefont{Y.-Y.} \bibnamefont{Wang}},
  \bibinfo{author}{\bibfnamefont{Z.-H.} \bibnamefont{Sun}}, \bibnamefont{and}
  \bibinfo{author}{\bibfnamefont{H.}~\bibnamefont{Fan}},
  \bibinfo{journal}{Phys. Rev. B} \textbf{\bibinfo{volume}{104}},
  \bibinfo{pages}{205122} (\bibinfo{year}{2021}).

\bibitem[{\citenamefont{Zhang et~al.}(2021)\citenamefont{Zhang, Ke, Liu, and
  Lee}}]{zhang2021mobility}
\bibinfo{author}{\bibfnamefont{L.}~\bibnamefont{Zhang}},
  \bibinfo{author}{\bibfnamefont{Y.}~\bibnamefont{Ke}},
  \bibinfo{author}{\bibfnamefont{W.}~\bibnamefont{Liu}}, \bibnamefont{and}
  \bibinfo{author}{\bibfnamefont{C.}~\bibnamefont{Lee}},
  \bibinfo{journal}{Phys. Rev. A} \textbf{\bibinfo{volume}{103}},
  \bibinfo{pages}{023323} (\bibinfo{year}{2021}).

\bibitem[{\citenamefont{Guo et~al.}(2021)\citenamefont{Guo, Cheng, Li, Xu,
  Zhang, Wang, Song, Liu, Ren, Dong et~al.}}]{guo2021stark}
\bibinfo{author}{\bibfnamefont{Q.}~\bibnamefont{Guo}},
  \bibinfo{author}{\bibfnamefont{C.}~\bibnamefont{Cheng}},
  \bibinfo{author}{\bibfnamefont{H.}~\bibnamefont{Li}},
  \bibinfo{author}{\bibfnamefont{S.}~\bibnamefont{Xu}},
  \bibinfo{author}{\bibfnamefont{P.}~\bibnamefont{Zhang}},
  \bibinfo{author}{\bibfnamefont{Z.}~\bibnamefont{Wang}},
  \bibinfo{author}{\bibfnamefont{C.}~\bibnamefont{Song}},
  \bibinfo{author}{\bibfnamefont{W.}~\bibnamefont{Liu}},
  \bibinfo{author}{\bibfnamefont{W.}~\bibnamefont{Ren}},
  \bibinfo{author}{\bibfnamefont{H.}~\bibnamefont{Dong}}, \bibnamefont{$et~al.$},
  \bibinfo{journal}{Phys. Rev. Lett.} \textbf{\bibinfo{volume}{127}},
  \bibinfo{pages}{240502} (\bibinfo{year}{2021}).

\bibitem[{\citenamefont{Yao et~al.}(2021)\citenamefont{Yao, Chanda, and
  Zakrzewski}}]{yao2021many}
\bibinfo{author}{\bibfnamefont{R.}~\bibnamefont{Yao}},
  \bibinfo{author}{\bibfnamefont{T.}~\bibnamefont{Chanda}}, \bibnamefont{and}
  \bibinfo{author}{\bibfnamefont{J.}~\bibnamefont{Zakrzewski}},
  \bibinfo{journal}{Phys. Rev. B} \textbf{\bibinfo{volume}{104}},
  \bibinfo{pages}{014201} (\bibinfo{year}{2021}).

\bibitem[{\citenamefont{Doggen et~al.}(2022)\citenamefont{Doggen, Gornyi, and
  Polyakov}}]{doggen2022many}
\bibinfo{author}{\bibfnamefont{E.~V.} \bibnamefont{Doggen}},
  \bibinfo{author}{\bibfnamefont{I.~V.} \bibnamefont{Gornyi}},
  \bibnamefont{and} \bibinfo{author}{\bibfnamefont{D.~G.}
  \bibnamefont{Polyakov}}, \bibinfo{journal}{Phys. Rev. B}
  \textbf{\bibinfo{volume}{105}}, \bibinfo{pages}{134204}
  (\bibinfo{year}{2022}).

\bibitem[{\citenamefont{Zisling et~al.}(2022)\citenamefont{Zisling, Kennes, and
  Lev}}]{zisling2022transport}
\bibinfo{author}{\bibfnamefont{G.}~\bibnamefont{Zisling}},
  \bibinfo{author}{\bibfnamefont{D.~M.} \bibnamefont{Kennes}},
  \bibnamefont{and} \bibinfo{author}{\bibfnamefont{Y.~B.} \bibnamefont{Lev}},
  \bibinfo{journal}{Phys. Rev. B} \textbf{\bibinfo{volume}{105}},
  \bibinfo{pages}{L140201} (\bibinfo{year}{2022}).

\bibitem[{\citenamefont{Burin}(2022)}]{burin2022exact}
\bibinfo{author}{\bibfnamefont{A.~L.} \bibnamefont{Burin}},
  \bibinfo{journal}{Phys. Rev. B} \textbf{\bibinfo{volume}{105}},
  \bibinfo{pages}{184206} (\bibinfo{year}{2022}).

\bibitem[{\citenamefont{Bertoni et~al.}(2022)\citenamefont{Bertoni, Eisert,
  Kshetrimayum, Nietner, and Thomson}}]{bertoni2022local}
\bibinfo{author}{\bibfnamefont{C.}~\bibnamefont{Bertoni}},
  \bibinfo{author}{\bibfnamefont{J.}~\bibnamefont{Eisert}},
  \bibinfo{author}{\bibfnamefont{A.}~\bibnamefont{Kshetrimayum}},
  \bibinfo{author}{\bibfnamefont{A.}~\bibnamefont{Nietner}}, \bibnamefont{and}
  \bibinfo{author}{\bibfnamefont{S.}~\bibnamefont{Thomson}},
  \bibinfo{journal}{arXiv:2208.14432}.

\bibitem[{\citenamefont{Lukin et~al.}(2022)\citenamefont{Lukin, Slyusarenko,
  and Sotnikov}}]{lukin2022many}
\bibinfo{author}{\bibfnamefont{I.}~\bibnamefont{Lukin}},
  \bibinfo{author}{\bibfnamefont{Y.~V.} \bibnamefont{Slyusarenko}},
  \bibnamefont{and} \bibinfo{author}{\bibfnamefont{A.}~\bibnamefont{Sotnikov}},
  \bibinfo{journal}{Phys. Rev. B} \textbf{\bibinfo{volume}{105}},
  \bibinfo{pages}{184307} (\bibinfo{year}{2022}).

\bibitem[{\citenamefont{Vernek}(2022)}]{vernek2022robustness}
\bibinfo{author}{\bibfnamefont{E.}~\bibnamefont{Vernek}},
  \bibinfo{journal}{Phys. Rev. B} \textbf{\bibinfo{volume}{105}},
  \bibinfo{pages}{075124} (\bibinfo{year}{2022}).

\bibitem[{\citenamefont{Pedersen et~al.}(2022)\citenamefont{Pedersen, Cornean,
  Krej{\v{c}}i{\v{r}}ik, Raymond, and Stockmeyer}}]{pedersen2022stark}
\bibinfo{author}{\bibfnamefont{T.}~\bibnamefont{Pedersen}},
  \bibinfo{author}{\bibfnamefont{H.}~\bibnamefont{Cornean}},
  \bibinfo{author}{\bibfnamefont{D.}~\bibnamefont{Krej{\v{c}}i{\v{r}}ik}},
  \bibinfo{author}{\bibfnamefont{N.}~\bibnamefont{Raymond}}, \bibnamefont{and}
  \bibinfo{author}{\bibfnamefont{E.}~\bibnamefont{Stockmeyer}},
  \bibinfo{journal}{New J. Phys.} \textbf{\bibinfo{volume}{24}},
  \bibinfo{pages}{093005} (\bibinfo{year}{2022}).

\bibitem[{\citenamefont{Sarkar and Buca}(2022)}]{sarkar2022protecting}
\bibinfo{author}{\bibfnamefont{S.}~\bibnamefont{Sarkar}} \bibnamefont{and}
  \bibinfo{author}{\bibfnamefont{B.}~\bibnamefont{Buca}},
  \bibinfo{journal}{arXiv:2204.13354}.

\bibitem[{\citenamefont{Yang et~al.}(2020)\citenamefont{Yang, Sun, Ott, Wang,
  Zache, Halimeh, Yuan, Hauke, and Pan}}]{yang2020observation}
\bibinfo{author}{\bibfnamefont{B.}~\bibnamefont{Yang}},
  \bibinfo{author}{\bibfnamefont{H.}~\bibnamefont{Sun}},
  \bibinfo{author}{\bibfnamefont{R.}~\bibnamefont{Ott}},
  \bibinfo{author}{\bibfnamefont{H.-Y.} \bibnamefont{Wang}},
  \bibinfo{author}{\bibfnamefont{T.~V.} \bibnamefont{Zache}},
  \bibinfo{author}{\bibfnamefont{J.~C.} \bibnamefont{Halimeh}},
  \bibinfo{author}{\bibfnamefont{Z.-S.} \bibnamefont{Yuan}},
  \bibinfo{author}{\bibfnamefont{P.}~\bibnamefont{Hauke}}, \bibnamefont{and}
  \bibinfo{author}{\bibfnamefont{J.-W.} \bibnamefont{Pan}},
  \bibinfo{journal}{Nature (London)} \textbf{\bibinfo{volume}{587}},
  \bibinfo{pages}{392} (\bibinfo{year}{2020}).

\bibitem[{\citenamefont{Halimeh et~al.}(2021)\citenamefont{Halimeh, Lang,
  Mildenberger, Jiang, and Hauke}}]{halimeh2021gauge}
\bibinfo{author}{\bibfnamefont{J.~C.} \bibnamefont{Halimeh}},
  \bibinfo{author}{\bibfnamefont{H.}~\bibnamefont{Lang}},
  \bibinfo{author}{\bibfnamefont{J.}~\bibnamefont{Mildenberger}},
  \bibinfo{author}{\bibfnamefont{Z.}~\bibnamefont{Jiang}}, \bibnamefont{and}
  \bibinfo{author}{\bibfnamefont{P.}~\bibnamefont{Hauke}},
  \bibinfo{journal}{PRX Quantum} \textbf{\bibinfo{volume}{2}},
  \bibinfo{pages}{040311} (\bibinfo{year}{2021}).

\bibitem[{\citenamefont{Zhou et~al.}(2022)\citenamefont{Zhou, Su, Halimeh, Ott,
  Sun, Hauke, Yang, Yuan, Berges, and Pan}}]{zhou2022thermalization}
\bibinfo{author}{\bibfnamefont{Z.-Y.} \bibnamefont{Zhou}},
  \bibinfo{author}{\bibfnamefont{G.-X.} \bibnamefont{Su}},
  \bibinfo{author}{\bibfnamefont{J.~C.} \bibnamefont{Halimeh}},
  \bibinfo{author}{\bibfnamefont{R.}~\bibnamefont{Ott}},
  \bibinfo{author}{\bibfnamefont{H.}~\bibnamefont{Sun}},
  \bibinfo{author}{\bibfnamefont{P.}~\bibnamefont{Hauke}},
  \bibinfo{author}{\bibfnamefont{B.}~\bibnamefont{Yang}},
  \bibinfo{author}{\bibfnamefont{Z.-S.} \bibnamefont{Yuan}},
  \bibinfo{author}{\bibfnamefont{J.}~\bibnamefont{Berges}}, \bibnamefont{and}
  \bibinfo{author}{\bibfnamefont{J.-W.} \bibnamefont{Pan}},
  \bibinfo{journal}{Science} \textbf{\bibinfo{volume}{377}},
  \bibinfo{pages}{311} (\bibinfo{year}{2022}).

\bibitem[{\citenamefont{Khemani et~al.}(2020)\citenamefont{Khemani, Hermele,
  and Nandkishore}}]{khemani2020localization}
\bibinfo{author}{\bibfnamefont{V.}~\bibnamefont{Khemani}},
  \bibinfo{author}{\bibfnamefont{M.}~\bibnamefont{Hermele}}, \bibnamefont{and}
  \bibinfo{author}{\bibfnamefont{R.}~\bibnamefont{Nandkishore}},
  \bibinfo{journal}{Phys. Rev. B} \textbf{\bibinfo{volume}{101}},
  \bibinfo{pages}{174204} (\bibinfo{year}{2020}).

\bibitem[{\citenamefont{Kohlert et~al.}(2021)\citenamefont{Kohlert, Scherg,
  Sala, Pollmann, Madhusudhana, Bloch, and
  Aidelsburger}}]{kohlert2021experimental}
\bibinfo{author}{\bibfnamefont{T.}~\bibnamefont{Kohlert}},
  \bibinfo{author}{\bibfnamefont{S.}~\bibnamefont{Scherg}},
  \bibinfo{author}{\bibfnamefont{P.}~\bibnamefont{Sala}},
  \bibinfo{author}{\bibfnamefont{F.}~\bibnamefont{Pollmann}},
  \bibinfo{author}{\bibfnamefont{B.~H.} \bibnamefont{Madhusudhana}},
  \bibinfo{author}{\bibfnamefont{I.}~\bibnamefont{Bloch}}, \bibnamefont{and}
  \bibinfo{author}{\bibfnamefont{M.}~\bibnamefont{Aidelsburger}},
  \bibinfo{journal}{arXiv:2106.15586}.

\bibitem[{\citenamefont{Scherg et~al.}(2021)\citenamefont{Scherg, Kohlert,
  Sala, Pollmann, Hebbe~Madhusudhana, Bloch, and
  Aidelsburger}}]{scherg2021observing}
\bibinfo{author}{\bibfnamefont{S.}~\bibnamefont{Scherg}},
  \bibinfo{author}{\bibfnamefont{T.}~\bibnamefont{Kohlert}},
  \bibinfo{author}{\bibfnamefont{P.}~\bibnamefont{Sala}},
  \bibinfo{author}{\bibfnamefont{F.}~\bibnamefont{Pollmann}},
  \bibinfo{author}{\bibfnamefont{B.}~\bibnamefont{Hebbe~Madhusudhana}},
  \bibinfo{author}{\bibfnamefont{I.}~\bibnamefont{Bloch}}, \bibnamefont{and}
  \bibinfo{author}{\bibfnamefont{M.}~\bibnamefont{Aidelsburger}},
  \bibinfo{journal}{Nat. Commun.} \textbf{\bibinfo{volume}{12}},
  \bibinfo{pages}{4490} (\bibinfo{year}{2021}).

\bibitem[{\citenamefont{Doggen et~al.}(2021)\citenamefont{Doggen, Gornyi, and
  Polyakov}}]{doggen2021stark}
\bibinfo{author}{\bibfnamefont{E.~V.} \bibnamefont{Doggen}},
  \bibinfo{author}{\bibfnamefont{I.~V.} \bibnamefont{Gornyi}},
  \bibnamefont{and} \bibinfo{author}{\bibfnamefont{D.~G.}
  \bibnamefont{Polyakov}}, \bibinfo{journal}{Phys. Rev. B}
  \textbf{\bibinfo{volume}{103}}, \bibinfo{pages}{L100202}
  (\bibinfo{year}{2021}).

\bibitem[{\citenamefont{Su et~al.}(2022)\citenamefont{Su, Sun, Hudomal,
  Desaules, Zhou, Yang, Halimeh, Yuan, Papi{\'c}, and Pan}}]{su2022observation}
\bibinfo{author}{\bibfnamefont{G.-X.} \bibnamefont{Su}},
  \bibinfo{author}{\bibfnamefont{H.}~\bibnamefont{Sun}},
  \bibinfo{author}{\bibfnamefont{A.}~\bibnamefont{Hudomal}},
  \bibinfo{author}{\bibfnamefont{J.-Y.} \bibnamefont{Desaules}},
  \bibinfo{author}{\bibfnamefont{Z.-Y.} \bibnamefont{Zhou}},
  \bibinfo{author}{\bibfnamefont{B.}~\bibnamefont{Yang}},
  \bibinfo{author}{\bibfnamefont{J.~C.} \bibnamefont{Halimeh}},
  \bibinfo{author}{\bibfnamefont{Z.-S.} \bibnamefont{Yuan}},
  \bibinfo{author}{\bibfnamefont{Z.}~\bibnamefont{Papi{\'c}}},
  \bibnamefont{and} \bibinfo{author}{\bibfnamefont{J.-W.} \bibnamefont{Pan}},
  \bibinfo{journal}{Phys. Rev. Res.}  \textbf{\bibinfo{volume}{5}},
  \bibinfo{pages}{023010} (\bibinfo{year}{2023}).

\bibitem[{\citenamefont{Morong et~al.}(2021)\citenamefont{Morong, Liu, Becker,
  Collins, Feng, Kyprianidis, Pagano, You, Gorshkov, and
  Monroe}}]{morong2021observation}
\bibinfo{author}{\bibfnamefont{W.}~\bibnamefont{Morong}},
  \bibinfo{author}{\bibfnamefont{F.}~\bibnamefont{Liu}},
  \bibinfo{author}{\bibfnamefont{P.}~\bibnamefont{Becker}},
  \bibinfo{author}{\bibfnamefont{K.}~\bibnamefont{Collins}},
  \bibinfo{author}{\bibfnamefont{L.}~\bibnamefont{Feng}},
  \bibinfo{author}{\bibfnamefont{A.}~\bibnamefont{Kyprianidis}},
  \bibinfo{author}{\bibfnamefont{G.}~\bibnamefont{Pagano}},
  \bibinfo{author}{\bibfnamefont{T.}~\bibnamefont{You}},
  \bibinfo{author}{\bibfnamefont{A.}~\bibnamefont{Gorshkov}}, \bibnamefont{and}
  \bibinfo{author}{\bibfnamefont{C.}~\bibnamefont{Monroe}},
  \bibinfo{journal}{Nature (London)} \textbf{\bibinfo{volume}{599}},
  \bibinfo{pages}{393} (\bibinfo{year}{2021}).

\bibitem[{\citenamefont{Preiss et~al.}(2015)\citenamefont{Preiss, Ma, Tai,
  Lukin, Rispoli, Zupancic, Lahini, Islam, and Greiner}}]{preiss2015strongly}
\bibinfo{author}{\bibfnamefont{P.~M.} \bibnamefont{Preiss}},
  \bibinfo{author}{\bibfnamefont{R.}~\bibnamefont{Ma}},
  \bibinfo{author}{\bibfnamefont{M.~E.} \bibnamefont{Tai}},
  \bibinfo{author}{\bibfnamefont{A.}~\bibnamefont{Lukin}},
  \bibinfo{author}{\bibfnamefont{M.}~\bibnamefont{Rispoli}},
  \bibinfo{author}{\bibfnamefont{P.}~\bibnamefont{Zupancic}},
  \bibinfo{author}{\bibfnamefont{Y.}~\bibnamefont{Lahini}},
  \bibinfo{author}{\bibfnamefont{R.}~\bibnamefont{Islam}}, \bibnamefont{and}
  \bibinfo{author}{\bibfnamefont{M.}~\bibnamefont{Greiner}},
  \bibinfo{journal}{Science} \textbf{\bibinfo{volume}{347}},
  \bibinfo{pages}{1229} (\bibinfo{year}{2015}).

\bibitem[{\citenamefont{Karamlou et~al.}(2022)\citenamefont{Karamlou,
  Braum{\"u}ller, Yanay, Di~Paolo, Harrington, Kannan, Kim, Kjaergaard,
  Melville, Muschinske et~al.}}]{karamlou2022quantum}
\bibinfo{author}{\bibfnamefont{A.~H.} \bibnamefont{Karamlou}},
  \bibinfo{author}{\bibfnamefont{J.}~\bibnamefont{Braum{\"u}ller}},
  \bibinfo{author}{\bibfnamefont{Y.}~\bibnamefont{Yanay}},
  \bibinfo{author}{\bibfnamefont{A.}~\bibnamefont{Di~Paolo}},
  \bibinfo{author}{\bibfnamefont{P.~M.} \bibnamefont{Harrington}},
  \bibinfo{author}{\bibfnamefont{B.}~\bibnamefont{Kannan}},
  \bibinfo{author}{\bibfnamefont{D.}~\bibnamefont{Kim}},
  \bibinfo{author}{\bibfnamefont{M.}~\bibnamefont{Kjaergaard}},
  \bibinfo{author}{\bibfnamefont{A.}~\bibnamefont{Melville}},
  \bibinfo{author}{\bibfnamefont{S.}~\bibnamefont{Muschinske}},
  \bibnamefont{$et~al.$}, \bibinfo{journal}{npj Quantum Inf.}
  \textbf{\bibinfo{volume}{8}}, \bibinfo{pages}{35} (\bibinfo{year}{2022}).

\bibitem[{\citenamefont{Fisher}(1922)}]{fisher1922mathematical}
\bibinfo{author}{\bibfnamefont{R.~A.} \bibnamefont{Fisher}},
  \bibinfo{journal}{Phil. Trans. R Soc. A} \textbf{\bibinfo{volume}{222}},
  \bibinfo{pages}{309} (\bibinfo{year}{1922}).

\bibitem[{\citenamefont{Meyer}(2021)}]{Meyer2021fisherinformationin}
\bibinfo{author}{\bibfnamefont{J.~J.} \bibnamefont{Meyer}},
  \bibinfo{journal}{{Quantum}} \textbf{\bibinfo{volume}{5}},
  \bibinfo{pages}{539} (\bibinfo{year}{2021}).

\bibitem[{\citenamefont{You et~al.}(2007)\citenamefont{You, Li, and
  Gu}}]{you2007fidelity}
\bibinfo{author}{\bibfnamefont{W.-L.} \bibnamefont{You}},
  \bibinfo{author}{\bibfnamefont{Y.-W.} \bibnamefont{Li}}, \bibnamefont{and}
  \bibinfo{author}{\bibfnamefont{S.-J.} \bibnamefont{Gu}},
  \bibinfo{journal}{Phys. Rev. E} \textbf{\bibinfo{volume}{76}},
  \bibinfo{pages}{022101} (\bibinfo{year}{2007}).

\bibitem[{\citenamefont{S.~Girvin}(2019)}]{girvin2019modern}
\bibinfo{author}{\bibfnamefont{S.~Girvin and K.~Yang} \bibnamefont{Rao}}, in
\emph{\bibinfo{booktitle}{Modern Condensed Matter Physics}}
(\bibinfo{publisher}{Cambridge University Press, Cambridge}, \bibinfo{year}{2019}), pp.
\bibinfo{pages}{252--300}.

\bibitem[{\citenamefont{Anderson}(1958)}]{anderson1958absence}
\bibinfo{author}{\bibfnamefont{P.~W.} \bibnamefont{Anderson}},
\bibinfo{journal}{Phys. Rev.} \textbf{\bibinfo{volume}{109}},
\bibinfo{pages}{1492} (\bibinfo{year}{1958}).

\bibitem[{\citenamefont{Luitz et~al.}(2015)\citenamefont{Luitz, Laflorencie,
  and Alet}}]{luitz2015many}
\bibinfo{author}{\bibfnamefont{D.~J.} \bibnamefont{Luitz}},
  \bibinfo{author}{\bibfnamefont{N.}~\bibnamefont{Laflorencie}},
  \bibnamefont{and} \bibinfo{author}{\bibfnamefont{F.}~\bibnamefont{Alet}},
  \bibinfo{journal}{Phys. Rev. B} \textbf{\bibinfo{volume}{91}},
  \bibinfo{pages}{081103(R)} (\bibinfo{year}{2015}).

\bibitem[{\citenamefont{Sorge}(2015)}]{andreas_sorge_2015_35293}
\bibinfo{author}{\bibfnamefont{A.}~\bibnamefont{Sorge}},
  \emph{\bibinfo{title}{\rm PYFSSA 0.7.6}} (\bibinfo{year}{2015}),
  \urlprefix\url{https://doi.org/10.5281/zenodo.35293}.

\bibitem[{\citenamefont{Melchert}(2009)}]{melchert2009autoscale}
\bibinfo{author}{\bibfnamefont{O.}~\bibnamefont{Melchert}},
  \bibinfo{journal}{arXiv:0910.5403}.
  
\bibitem[{\citenamefont{Hauschild and Pollmann}(2018)}]{TENPY}
\bibinfo{author}{\bibfnamefont{J.}~\bibnamefont{Hauschild}} \bibnamefont{and}
  \bibinfo{author}{\bibfnamefont{F.}~\bibnamefont{Pollmann}},
  \bibinfo{journal}{SciPost Phys. Lecture Notes} ~\bibinfo{pages}{5}
  (\bibinfo{year}{2018}), \bibinfo{note}{code available from
  \url{https://github.com/tenpy/tenpy}}, \eprint{1805.00055},
  \urlprefix\url{https://scipost.org/10.21468/SciPostPhysLectNotes.5}.
  

\end{thebibliography}


\clearpage
\widetext
\setcounter{equation}{0}
\setcounter{figure}{0}
\setcounter{table}{0}
\setcounter{page}{1}
\setcounter{section}{0}
\makeatletter
\renewcommand{\theequation}{S\arabic{equation}}
\renewcommand{\thefigure}{S\arabic{figure}}
\renewcommand{\thetable}{S\arabic{table}}
\renewcommand{\bibnumfmt}[1]{[S#1]}
\renewcommand{\citenumfont}[1]{S#1}

\begin{center}
\textbf{\large Supplemental Material to "Stark localization as a resource for weak-field sensing with super-Heisenberg precision"}
\end{center}

\author{Xingjian He}%
\affiliation{Institute of Fundamental and Frontier Sciences, University of Electronic Science and Technology of China, Chengdu 610051, China}

\author{Rozhin Yousefjani}%
\affiliation{Institute of Fundamental and Frontier Sciences, University of Electronic Science and Technology of China, Chengdu 610051, China}

\author{Abolfazl Bayat}%
\affiliation{Institute of Fundamental and Frontier Sciences, University of Electronic Science and Technology of China, Chengdu 610051, China}


\section{Saturation of QFI in Stark localized regime}
\begin{figure}[h!]
\includegraphics[width=0.65\linewidth]{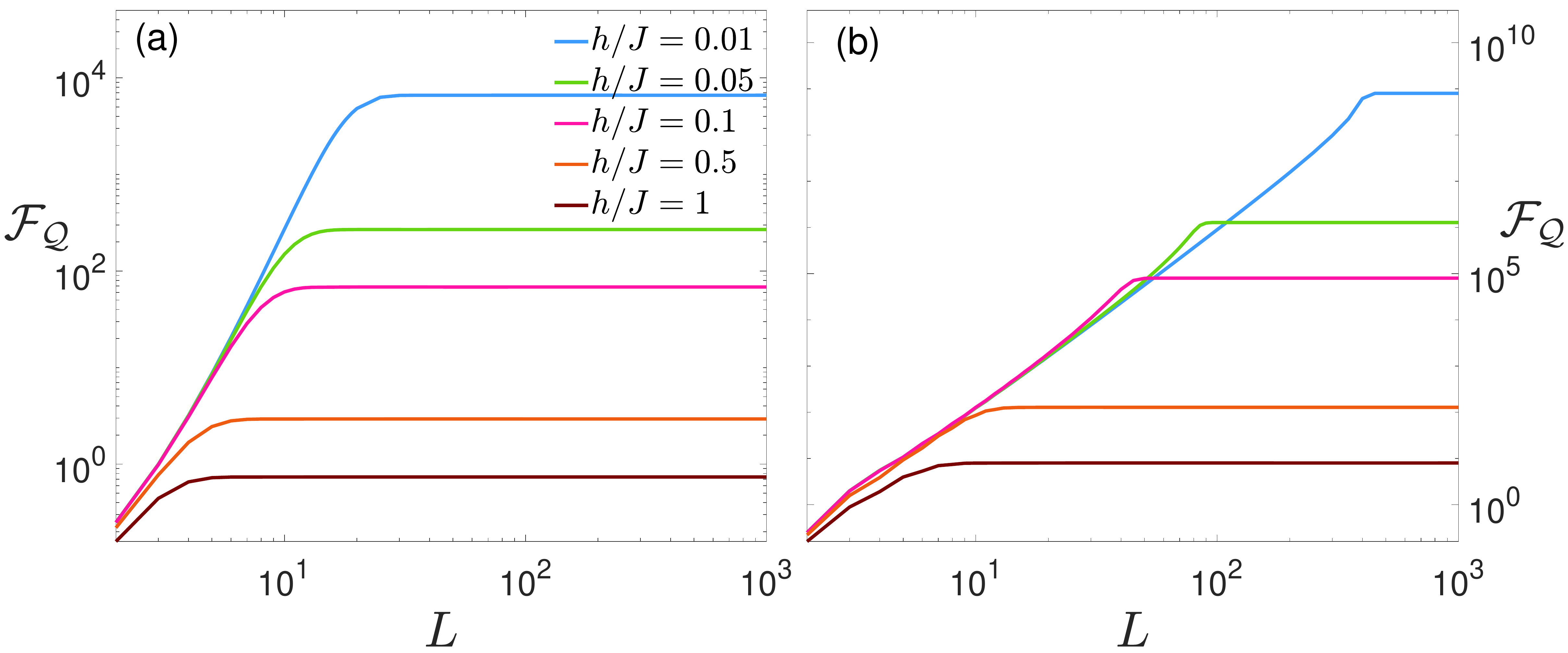} 
\caption{In the localized phase, with various choices of $h/J$, the saturation of the QFI with respect to $L$ is depicted for: (a) the ground state; and (b) the midspectrum eigenstate. }\label{fig:FigS1}
\end{figure}

Followed by our discussion in the main text about the algebraic behavior of QFI, in this section we present numerical evidence to show its saturation with size in the localized phase. In Figs.~\ref{fig:FigS1}(a) and \ref{fig:FigS1}(b) we plot $\mathcal{F}_Q(h)$ as a function of system size $L$ for various choices of gradient field $h/J$ for both ground and midspectrum eigenstates, respectively. 
By deviating from the transition point $h_{\max}$ the convergence becomes faster and QFI saturates to its thermodynamic value, which indicates that one can emulate thermodynamic behavior of Stark probe using finite systems.

\section{Whole-spectrum criticality}
Following the main text, we establish a close relationship, as Eq.~(4), among the three critical exponents governing Stark localization transition for both ground and midspectrum eigenstates. In this section we show that this relationship holds for all the eigenstates across the spectrum through extensive finite-size scaling analysis.
In TABLE.~\ref{TABLE:exponent} we report the obtained values of $\alpha$, $\nu$, $\alpha/\nu$ and $\beta$ across the whole spectrum. Three important features can be observed: (i) the critical exponents found for the ground state are quite different from the ones obtained for other values of energy; (ii) one can observe a symmetry between critical exponents of the different energies, the same symmetry has been observed in Fig.~1(c) and (d) in the main text; and (iii) Eq.~(4) remains valid with high accuracy across the whole spectrum, signaling the generality of our analysis.

\begin{table}[h!]
\resizebox{0.9\linewidth}{!}{
\setlength{\tabcolsep}{3mm}{
\renewcommand{\arraystretch}{1.3}
\begin{tabular}{|c|c|c|c|c|c|c|c|c|c|c|c|}
\specialrule{0em}{3pt}{3pt}
\hline
$\varepsilon$   & 0     & 0.1   & 0.2   & 0.3   & 0.4   & 0.5   & 0.6   & 0.7   & 0.8   & 0.9   & 1     \\ \hline
$\alpha$   & 2.00 & 3.75 & 3.90 & 3.97 & 4.00 & 4.00 & 4.00 & 3.97 & 3.90 & 3.75 & 2.00 \\ \hline
$\nu$   & 0.33 & 0.94 & 0.98 & 1.00 & 1.00 & 1.00 & 1.00 & 1.00 & 0.98 & 0.94 & 0.33 \\ \hline
$\alpha/\nu$ & 5.98 & 3.97 & 3.98 & 3.99 & 3.99 & 3.99 & 3.99 & 3.99 & 3.98 & 3.97 & 5.98 \\ \hline
$\beta$   & 5.98 & 4.13 & 4.09 & 4.07 & 4.06 & 4.11 & 4.06 & 4.07 & 4.09 & 4.13 & 5.98 \\ \hline
\end{tabular}
 }
 }
 \caption{The critical exponents across the spectrum. The critical exponents $\alpha$ and $\nu$ are obtained from the finite-size scaling analysis in Fig.~3 while the exponent $\beta$ is extracted from the data analysis in Fig.~2.  }
\label{TABLE:exponent}
\end{table}

\section{Accessibility of super-Heisenberg scaling}
\begin{figure}[h!]
\includegraphics[width=0.49\linewidth]{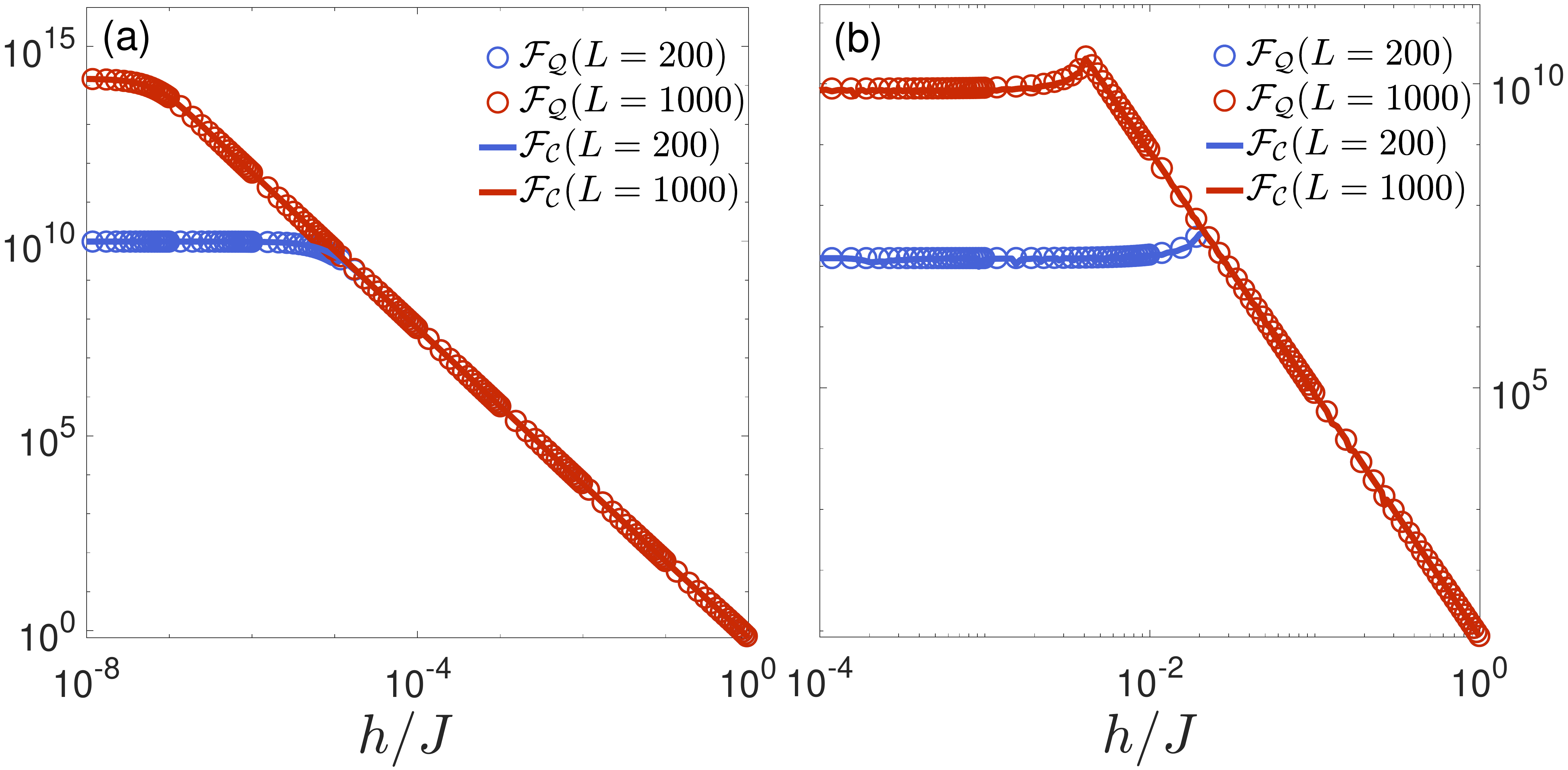}
\includegraphics[width=0.49\linewidth]{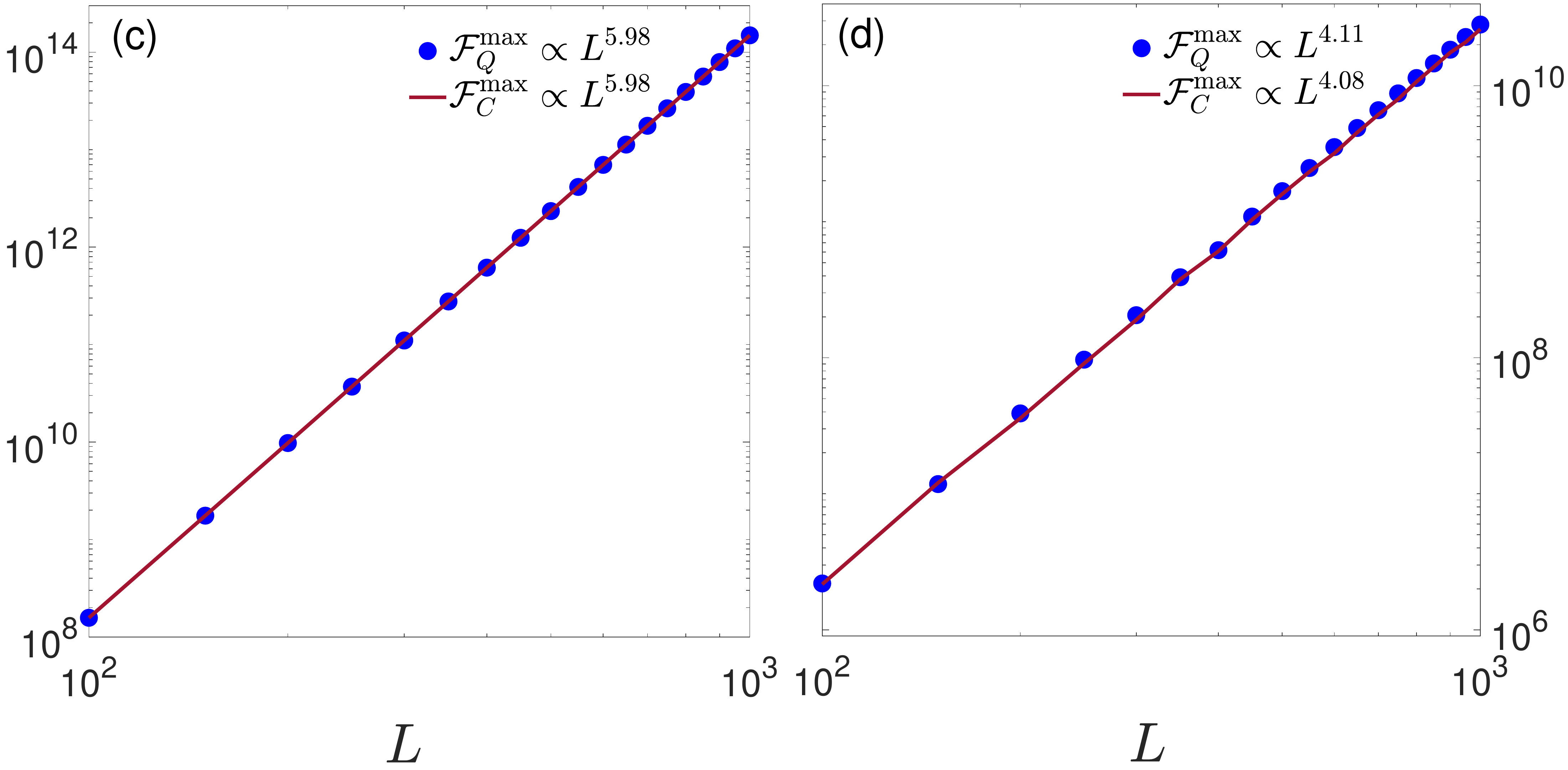}
\caption{The CFI and QFI as a function of Stark field $h/J$ when our probe with system sizes $L=200,1000$ is prepared in (a) the ground state and (b) the midspectrum eigenstate. The maximum of QFI (markers) and CFI (solid lines) as a function of system size $L$ for (c) the ground state and (d) the midspectrum eigenstate. The optimal data fitting for numerical results of CFI is obtained as $\mathcal{F}_{C}(h_{\max})\propto L^{\beta}$ with $\beta=5.98$ and $\beta=4.08$ evaluated for the lowest and midspectrum energy levels, respectively.
}\label{fig:FigS2}
\end{figure}

In the main text of the letter, we present a simple position measurement described
by local projective operators as 
$\{\Pi_{i}=\vert i\rangle\langle i\vert\}_{i=1}^{L}$ to nearly saturate the quantum Cram\'{e}r-Rao bound.
This section deals with providing numerical evidence for this claim.
For $p_i(h)=\mathrm{Tr}[\Pi_{i}\rho(h)]$ as the probability of finding the particle in site $i$, the CFI can be calculated using 
$\mathcal{F}_{C}(h)=\sum_{i}p_{i}(h)[\partial_{h}\ln p_{i}(h)]^2$.
In Figs.~\ref{fig:FigS2}(a) and \ref{fig:FigS2}(b), $\mathcal{F}_{Q}$ (markers) and $\mathcal{F}_{C}$ (solid line) as a function of the gradient field $h/J$  are plotted for two different probe sizes prepared in the ground state and midspectrum eigenstate, respectively.
Surprisingly, the CFI obtained via position measurement highly resembles that of QFI for both scenarios.
The maximum (peak) of CFI happens in the same $h_{\max}$ for the QFI and by increasing the system size from $L=200$ to $L=1000$ smoothly skews toward vanishingly smaller values of $h$.
Similar to the QFI, the maximum of the CFI significantly increases, signaling the divergence of the CFI in the thermodynamic limit.
The divergent and size-independent behaviors of the CFI in both extended and localized regimes hint that analog to the QFI, the thermodynamic behavior of the system far from criticality can be emulated by finite-size systems that are measured by $\{\Pi_{i}=\vert i\rangle\langle i\vert\}_{i=1}^{L}$.
This is a powerful witness of closely saturating the quantum Cram\'{e}r-Rao bound. To capture the scaling behavior of the CFI with respect to probe size $L$, in Figs.~\ref{fig:FigS2}(c) and \ref{fig:FigS2}(d) we plot $\mathcal{F}_{C}(h_{\max})$ (solid line) as well as  $\mathcal{F}_{Q}(h_{\max})$ (marker)
for both lowest and midspectrum energy levels, respectively.
Obviously, the CFI follows the QFI scaling behavior which can be described by $\mathcal{F}_{C}(h_{\max})\propto L^{\beta}$.
We found that for the ground state, the scaling exponent $\beta$ of CFI converges to that of QFI. While for the midspectrum state, both the value and the scaling exponent $\beta$ of QFI are quite close to those of CFI.
This numerical simulation shows that the simple position measurement in our Stark probe is the optimal measurement setup that can saturate the quantum Cram\'{e}r-Rao bound and achieve super-Heisenberg scaling.

\section{Algebraic behavior in thermal equilibrium when $KT{>}\Delta E$}
\begin{figure}[h!]
\includegraphics[width=0.65\linewidth]{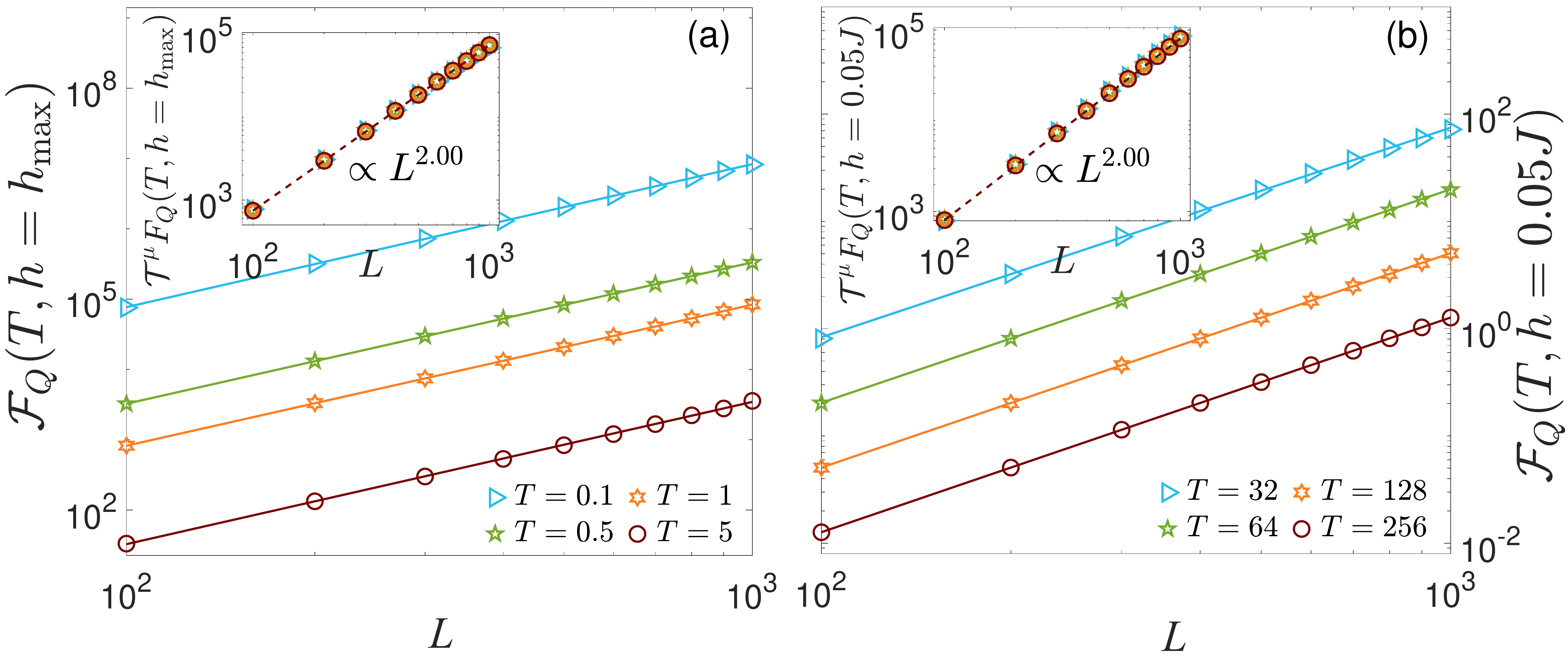}
\caption{The scaling behavior of QFI when $KT>\Delta E$  and the Stark probe operate in its (a) critical and (b) localized regimes. The rescaled QFI as a function of $L$ is plotted in the insets of (a) and (b). The best data collapse has been obtained as $T^{\mu}\mathcal{F}_{Q}\propto L^{\gamma}$ with $\gamma=2.00$ for both considered regimes.}
\label{fig:FigS3}
\end{figure}

In this section, we present numerical evidence to show the universal behavior, namely $\mathcal{F}_{Q}(h)\propto f(h)T^{-\mu}L^{\gamma}$, of the Stark probe in the thermodynamic equilibrium.
In Figs.~\ref{fig:FigS3}(a) and \ref{fig:FigS3}(b), the scaling behavior of the QFI with respect to probe size $L$ in different temperatures $T$ has been plotted for two scenarios.
In panel (a) the Stark probe operates in its critical regime (i.e. $h=h_{\max}$).
In contrast, in panel (b), the gradient field is considered large enough ($h=0.05J$) to assess the functionality of the probe in its off-critical regime.
Clearly, for both scenarios, increasing the temperature reduces the QFI. 
The universal scaling behavior of the QFI with respect to probe size can be obtained by establishing a finite-size scaling analysis. In the insets of Figs.~\ref{fig:FigS3}(a) and \ref{fig:FigS3}(b), we plot the rescaled QFI (i.e., $T^{\mu}\mathcal{F}_{Q}$) with respect to $L$ for critical ($h=h_{\max}$) and off-critical ($h=0.05J$) regimes, respectively.
For both scenarios, the best data collapse is obtained as $T^{\mu}\mathcal{F}_{Q}\propto L^{\gamma}$ with $\gamma=2.00$  for the critical and off-critical regimes. 
This numerical simulation guarantees the validity of the universal behavior of thermal probe as $\mathcal{F}_{Q}\propto T^{-2}L^{2}$.
Obviously, followed by the super-Heisenberg scaling with the ground state for $KT<\Delta E$ at $T\rightarrow0$, the Heisenberg precision can still be reachable for our Stark probe in a higher temperature and localized regime.

\section{Resource assessment in the adiabatic limit}

\begin{figure}[h!]
\includegraphics[width=0.49\linewidth]{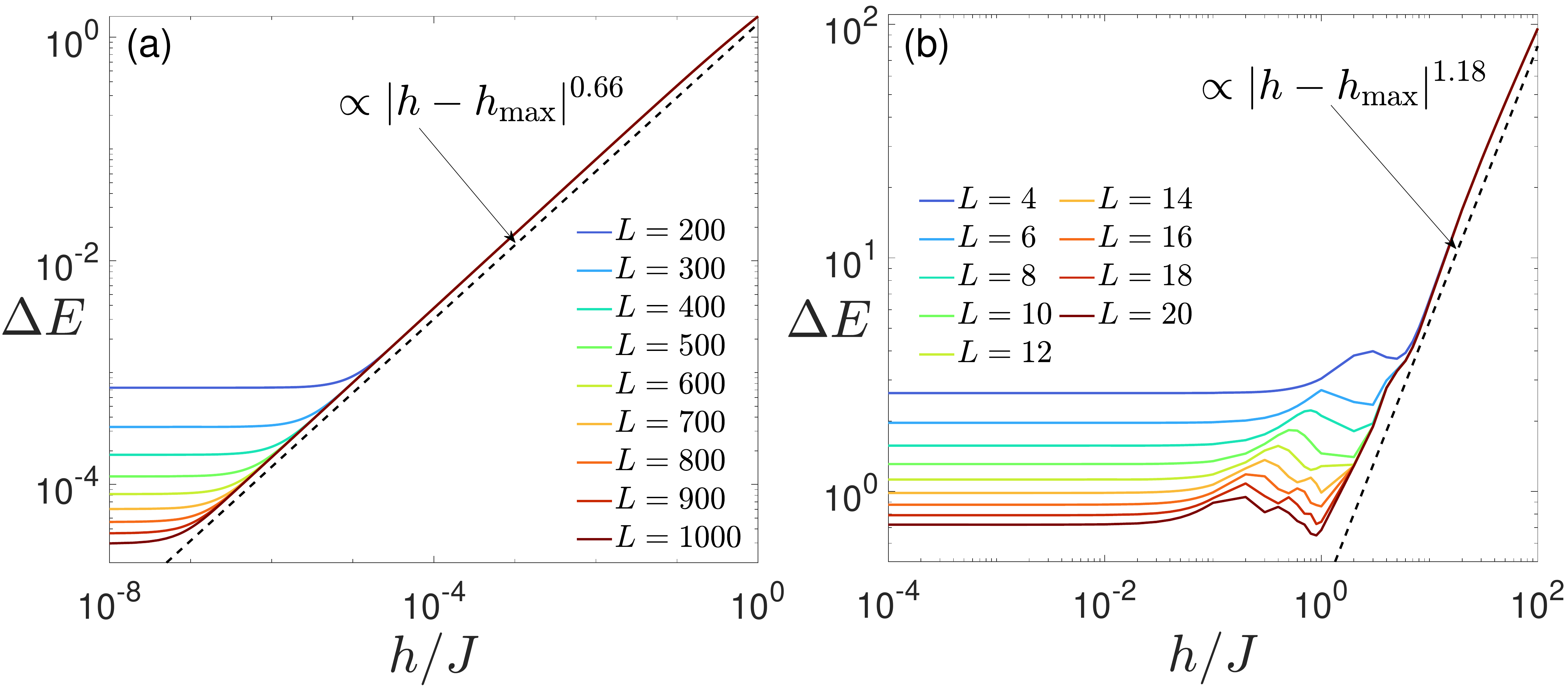}
\includegraphics[width=0.49\linewidth]{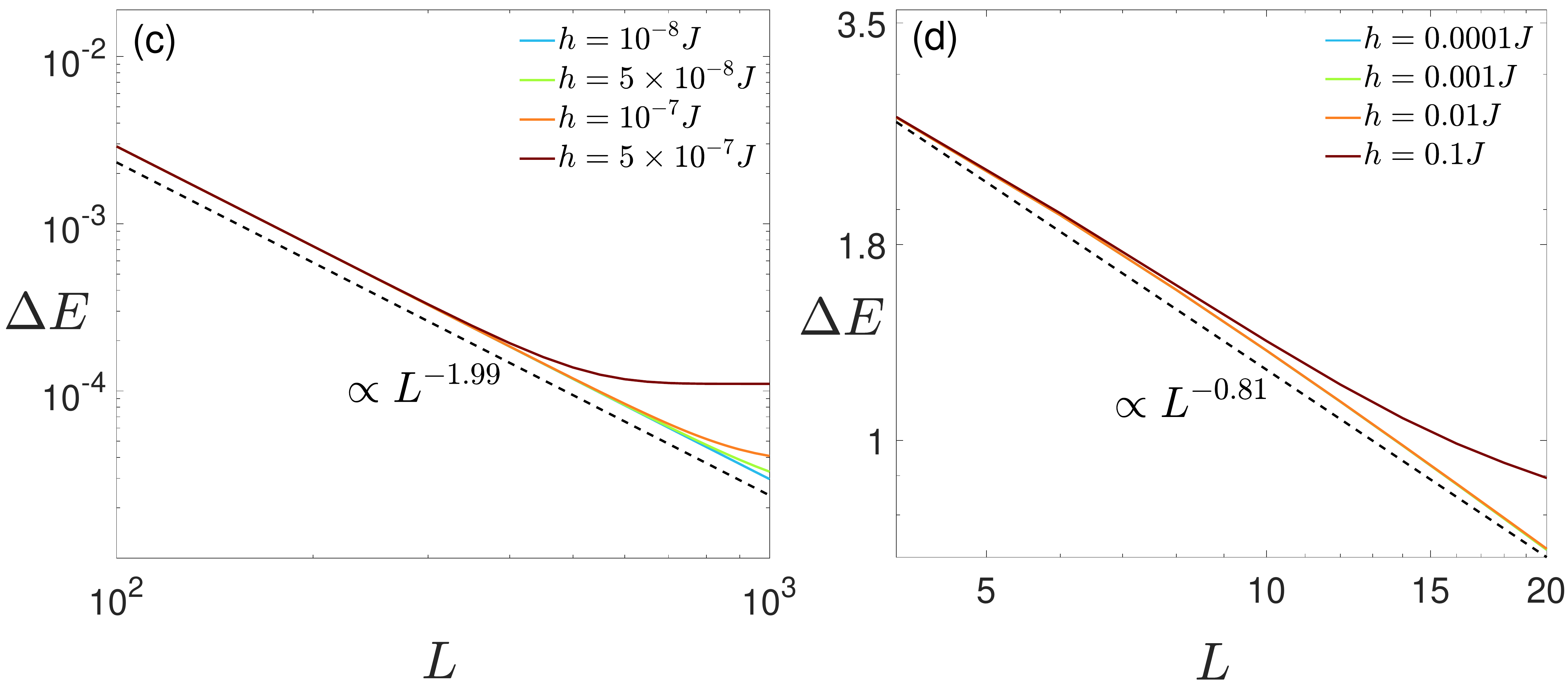}
\caption{The energy gap $\Delta E$ as a function of Stark field $h/J$ for different system sizes in: (a) single-particle probes; and (b) many-body interacting probes. The dashed lines in both panels show the algebraic behavior of $\Delta E$ in the localized phase. The energy gap $\Delta E$ as a function of system size $L$ for various values of $h/J$ is depicted for: (c) single-particle and (d) many-body probes.}
\label{fig:FigS4}
\end{figure}

Using an adiabatic evolution for ground state initialization near the critical point requires a preparation size-dependent time which scales as $t\sim L^{z}$, where $z$ is the dynamical critical exponent governing the speed of gap closing near the criticality, namely $\Delta E\sim L^{-z}$. To include the preparation time in  the resource analysis of the protocol, one can use normalized QFI $\mathcal{F}_Q/t$ as a figure of merit. Therefore, the scaling changes to $\mathcal{F}_Q/t\sim L^{\beta-z}$. In this section, we first estimate the dynamical critical exponent $z$ through analysis of $\Delta E$. 
In Figs.~\ref{fig:FigS4}(a) and \ref{fig:FigS4}(b), the energy gap $\Delta E$ as a function of Stark field $h/J$ is plotted for single-particle and many-body interacting probes, respectively. Energy gap $\Delta E$ takes its minimum value at the critical point $h=h_{\max}$. In Fig.~\ref{fig:FigS4}(a), the minimum takes place at vanishingly small fields, namely $h_{\max}\rightarrow0$, for the single-particle case. For many-body probes, see Fig.~\ref{fig:FigS4}(b), a clear drop of $\Delta E$ can be observed at the transition point. The dashed lines also indicate an algebraic behavior of the form $\Delta E\propto|h-h_{\max}|^{\eta}$ in the localized phase, which can be perfectly fitted by $\eta=0.66$ for the single-particle and $\eta=1.18$ for many-body probes. To see how the $\Delta E$ scales with the probe size, in Figs.~\ref{fig:FigS4}(c) and \ref{fig:FigS4}(d), we plot  $\Delta E$ versus $L$ in the extended phase as well as the vicinity of the critical point, respectively. The dynamical critical exponent $z$ extracted from the critical point is $z\simeq2$ for the single-particle and $z\simeq0.81$ for the many-body probe. Hence, our new figure of merit $\mathcal{F}_Q/t$ shows scaling as $\mathcal{F}_Q/t\sim L^{4}$ for single-particle and $\mathcal{F}_Q/t\sim L^{3.45}$ for many-body interacting probes. Remarkably, in both of these cases  quantum-enhanced sensitivity remains valid showing superiority over classical sensors.

\end{document}